\documentclass[sigconf]{acmart}
\renewcommand\footnotetextcopyrightpermission[1]{} 
\usepackage{setspace} 
\usepackage{booktabs} 
\usepackage{listings}
\usepackage{xcolor}
\usepackage{graphicx}
\usepackage{url}
\usepackage{amsmath}
\usepackage{booktabs}
\usepackage{multirow}
\usepackage{lscape}
\usepackage{xspace}
\usepackage[T1]{fontenc}
\usepackage[scaled=0.75]{beramono}
\usepackage{marvosym}
\usepackage{pifont}
\usepackage{float}
\usepackage{tikz}
\usepackage{comment}
\usepackage{enumerate}
\usepackage{array}
\usepackage{hyperref}
\usepackage{framed}
\usepackage{color}
\usepackage{enumitem}
\usepackage{subfigure}
\usepackage[ruled,linesnumbered]{algorithm2e}
\usepackage[normalem]{ulem} 
\definecolor{light-gray}{gray}{0.7}
\newcommand\hl{\bgroup\markoverwith
  {\textcolor{light-gray}{\rule[-.5ex]{2pt}{2.5ex}}}\ULon}

\newcommand{\tabincell}[2]{\begin{tabular}{@{}#1@{}}#2\end{tabular}}
\newcommand{\etal}{et.~al}
\usepackage{titlesec}
\titleformat*{\section}{\LARGE\bfseries}
\titleformat*{\subsubsection}{\large\bfseries\slshape}
\titlespacing*{\subsubsection}{0pt}{8pt}{5pt}

\begin{document}
\title{EVMFuzz: Differential Fuzz Testing of Ethereum Virtual Machine} 

\author{Ying Fu}
\affiliation{Tsinghua University, China}
\email{fy17@mails.tsinghua.edu.cn}

\author{Meng Ren}
\affiliation{Sun Yat-sen University, China}
\email{renm8@mail2.sysu.edu.cn}

\author{Fuchen Ma}
\affiliation{Beijing University of Posts and Telecommunications, China}
\email{mfc@bupt.edu.cn}

\author{Yu Jiang}
\authornote{Yu Jiang is the correspondence author.}
\affiliation{Tsinghua University, China}
\email{jiangyu198964@126.com}

\author{Heyuan Shi}
\affiliation{Tsinghua University, China}
\email{shy15@mails.tsinghua.edu.cn}

\author{Jiaguang Sun}
\affiliation{Tsinghua University, China}
\email{sunjiaguang@126.com}

\begin{abstract}
Ethereum Virtual Machine (EVM) is the run-time environment for smart contracts and its vulnerabilities may lead to serious problems to the Ethereum ecology. 
With lots of techniques being developed for the validation of smart contracts, the security problems of EVM have not been well-studied. 

In this paper, we propose EVMFuzz, aiming to detect vulnerabilities of EVMs with differential fuzz testing. 
The core idea of EVMFuzz is to continuously generate seed contracts for different EVMs' execution, so as to find as many inconsistencies among execution results as possible, eventually discover vulnerabilities with output cross-referencing. 
First, we present the evaluation metric for the internal inconsistency indicator, such as the opcode sequence executed and gas used. Then, we construct seed contracts via a set of predefined mutators and employ dynamic priority scheduling algorithm to guide seed contracts selection and maximize the inconsistency. 
Finally, we leverage different EVMs as cross-referencing oracles to avoid manual checking of the execution output. 
For evaluation, we conducted large-scale mutation on 36,295 real-world smart contracts and generated 253,153 smart contracts. Among them, 66.2\% showed differential performance, including 1,596 variant contracts triggered inconsistent output among EVMs. Accompanied by manual root cause analysis, we found 5 previously unknown security bugs in four widely used EVMs, and all had been included in Common Vulnerabilities and Exposures (CVE) database. 

\end{abstract}

\keywords{Differential testing, fuzzing, domain-specific mutation, EVM} 

\maketitle

\section{Introduction}
\label{introduction}
Ethereum can be viewed as a transaction-based state machine~\cite{Wood2014EthereumAS}, whose foundation is transaction execution. There are about 500,000 transactions~\cite{TxOnEtherscan} running on Ethereum every day, most of which involve the execution of smart contracts. 
Over the past few years, the safety and security problems of the blockchain transaction caused by smart contracts have emerged endlessly. In June 2016, Blockchain industry's largest crowdfunding project --- $TheDAO$, was attacked due to a serious flaw in $splitDAO$ function, resulting in more than three million Ether loss~\cite{dao}. Then, many efforts are devoted to safeguarding the transaction security by ensuring rigorous code logic of smart contracts. For instance, Oyente~\cite{Luu2016MakingSC} and MAIAN~\cite{Nikolic2018FindingTG} use symbolic execution techniques to find potential security vulnerabilities in Solidity smart contracts. Zeus~\cite{Kalra2018ZEUSAS} applies abstract interpretation to analyze smart contracts. 

However, as the authentic platform and standard for executing smart contracts, if there are some vulnerabilities in EVM's internal implementation, it will definitely lead to serious consequences such as Ethereum operation errors or transaction failures. 
We selected 10 contracts that passed Oyente~\cite{Luu2016MakingSC} tool examination and executed them on EVMs implemented by JavaScript, Python, C++ and Go language, finding that 7 of them had different gas consumption on different platforms at the end of execution, and one function even had the result of execution failure on Py-EVM~\cite{pyevm}, which is the Python implementation of the EVM. This shows that only contract validation cannot ensure the correctness of Ethereum transactions execution, and it is of great urgency to secure EVM. To perform efficient EVM testing, we need to solve the following challenges:

\begin{itemize}[leftmargin=*]
\item \textbf{How to define general scenarios and evaluation metrics? }
~\\ At present, EVM has at least 10 widely used implementations of different programming language~\cite{impl}, all of which are based on the standards of Ethereum Yellow Paper~\cite{Wood2014EthereumAS}. For example, the amount of EVM code in Geth~\cite{geth} platform is about 5,475 lines of Go, accounting for 27.6\% of the core infrastructure implementation of Geth; the data in Parity~\cite{parity} is about 18,912 lines of Rust, accounting for 31.7\% of the core infrastructure implementation. The implementation of EVM is complex, involving a large number of control structures and storage structures.
Because of the EVM version diversity and code complexity, it is necessary to define meaningful and general evaluation metric for the testing of different EVMs. 

\item \textbf{How to generate test cases that trigger EVM bugs? }
~\\ The Ethereum Yellow Paper provides the basis for all implementations of EVM, in which the functions and attributes are defined by formulas and rules. But there is neither authoritative test suites and benchmarks nor the common vulnerabilities and defects of EVM platforms sorted out by authority, which makes EVM testing loses the basis and target of detection. 
Moreover, there is no mature testing tool for EVM which makes it difficult to employ EVM testing on a large scale in the short term.
Due to the lack of official benchmarks and widely used EVM testing frameworks, we urgently need to find out a solution which is able to perform efficient and accurate EVM testing, automatically generating EVM inputs fast and effective.

\end{itemize}

To address these challenges, we implement EVMFuzz, which aims to automatically generate adversarial test inputs for EVMs. 

First, we define a general evaluation metric for the differential fuzzing of EVMs. As most EVMs are implemented as a transaction-based state machine, and the change of state depends on the opcode sequence to be executed, the input parameters and gas limit, 
hence, we use the opcode sequence executed and gas used as two important indicators to evaluate EVMs' performance on each test contract. 
EVMFuzz integrates different EVMs and creates a unified running environment for them. In this way, it takes the natural advantages of multiple versions to quickly discover the output inconsistencies without manual checking. Then, our seed contract mutation and selection algorithms can continuously generate contracts that enlarge the metric difference, so that EVMFuzz can efficiently mine cases that trigger differential performance of EVMs and try to get those corner cases with inconsistent execution output. 

For evaluation, we firstly conducted empirical studies on 36,295 real-world smart contracts from Etherscan~\cite{etherscan} and found that 24,000 contracts triggered metric inconsistencies among different EVMs, even with the same execution output. Through guided fuzzing, 
1,596 variant contracts successfully triggered inconsistent execution output among different EVMs. With manual analysis, we found 5 previously unknown security bugs in different widely used EVMs, and all had been included in Common Vulnerabilities and Exposures (CVE) database~\cite{cve}.

\vspace{3pt}
\noindent \emph{\textbf{Contributions}} \quad
We make the following main contributions:
\begin{itemize}[leftmargin=*]
\item We introduce an evaluation metric for EVM differential fuzzing, define 8 mutators for seed contract generation and design dynamic priority scheduling algorithm for seed contract selection.
\item We implement EVMFuzz, an automated differential fuzz testing framework, to efficiently expose the differences and vulnerabilities of different EVMs. 
\item We apply EVMFuzz to test some most widely used EVM versions, many inconsistencies and security bugs are detected, and 5 vulnerabilities have been assigned with unique CVE IDs.
\end{itemize}

\vspace{3pt}
\noindent \emph{\textbf{Paper Organization}} \quad
The rest of this paper is organized as follows.
Section \ref{background} introduces some background and gives a motivating example.
We provide a high-level overview of EVMFuzz in Section \ref{approach}, and the details implementation are described in Section \ref{Methodology}. 
The evaluation results are shown in Section \ref{Evaluation}. Section \ref{Discussion} outlines the limitations and proposes future directions for improvement.
Finally, we survey related work in Section \ref{Related-work} and conclude in Section \ref{Conclusion}.
\section{Background}
\label{background}

\subsection{The Ethereum Virtual Machine}
Ethereum Virtual Machine (EVM) is the heart of the Ethereum, which is often called the operating system of the Ethereum technology and is 
responsible for the execution and maintenance of smart contracts. It is the bedrock on which smart contracts are built.

The formal definition of the EVM is specified in the Ethereum Yellow Paper~\cite{Wood2014EthereumAS}. EVM is a simple stack-based architecture, whose word size (size of stack items) is 256-bit. According to the predefined execution environment and execution steps, such as exception halting and jump destination validity, it completes the state transition of each Ethereum block. EVM handles the execution of bytecode and the calculation of gas consumption. 
In general, EVM is a powerful, sandboxed virtual stack embedded within each full Ethereum node, responsible for executing contract bytecode. 

In accordance with the standards of Ethereum Yellow Paper, EVMs have been successfully implemented in various programming languages including C++, Go and many others~\cite{impl}.
There are tens of thousands of people doing transactions via the clients based on these EVM implementations everyday. 
Therefore, the vulnerability hidden in any EVM version might result in serious consequences.

\subsection{A Motivating Example}

To investigate the effectiveness of EVMFuzz, we use a simple example presented below. $forTest$ is a simple contract with a function $Testwhile$, whose data structure is a $while$ loop. When the input parameter $a$ is less than $b$, variable $x$ will continually increase. Strictly speaking, the implementation of $Testwhile$ function has a serious problem and may not appear in real life, if the input parameter satisfies the loop condition, it will result in an infinite loop. However, such contract can pass the check of most existing contract testing and verification tools, and our experiment proves that it can trigger different behaviors of multiple EVM implementations and even cause the denial of service problem. 


\definecolor{verylightgray}{rgb}{.97,.97,.97}

\lstdefinelanguage{Solidity}{
	keywords=[1]{anonymous, assembly, assert, balance, break, call, callcode, case, catch, class, constant, continue, constructor, contract, debugger, default, delegatecall, delete, do, else, emit, event, experimental, export, external, false, finally, for, function, gas, if, implements, import, in, indexed, instanceof, interface, internal, is, length, library, log0, log1, log2, log3, log4, memory, modifier, new, payable, pragma, private, protected, public, pure, push, require, return, returns, revert, selfdestruct, send, solidity, storage, struct, suicide, super, switch, then, this, throw, transfer, true, try, typeof, using, value, view, while, with, addmod, ecrecover, keccak256, mulmod, ripemd160, sha256, sha3}, 
	keywordstyle=[1]\color{blue}\bfseries,
	keywords=[2]{address, bool, byte, bytes, bytes1, bytes2, bytes3, bytes4, bytes5, bytes6, bytes7, bytes8, bytes9, bytes10, bytes11, bytes12, bytes13, bytes14, bytes15, bytes16, bytes17, bytes18, bytes19, bytes20, bytes21, bytes22, bytes23, bytes24, bytes25, bytes26, bytes27, bytes28, bytes29, bytes30, bytes31, bytes32, enum, int, int8, int16, int24, int32, int40, int48, int56, int64, int72, int80, int88, int96, int104, int112, int120, int128, int136, int144, int152, int160, int168, int176, int184, int192, int200, int208, int216, int224, int232, int240, int248, int256, mapping, string, uint, uint8, uint16, uint24, uint32, uint40, uint48, uint56, uint64, uint72, uint80, uint88, uint96, uint104, uint112, uint120, uint128, uint136, uint144, uint152, uint160, uint168, uint176, uint184, uint192, uint200, uint208, uint216, uint224, uint232, uint240, uint248, uint256, var, void, ether, finney, szabo, wei, days, hours, minutes, seconds, weeks, years},	
	keywordstyle=[2]\color{teal}\bfseries,
	keywords=[3]{block, blockhash, coinbase, difficulty, gaslimit, number, timestamp, msg, data, gas, sender, sig, value, now, tx, gasprice, origin},	
	keywordstyle=[3]\color{violet}\bfseries,
	identifierstyle=\color{black},
	sensitive=false,
	comment=[l]{//},
	morecomment=[s]{/*}{*/},
	commentstyle=\color{gray}\ttfamily,
	stringstyle=\color{red}\ttfamily,
	morestring=[b]',
	morestring=[b]"
}

\lstset{
	language=Solidity,
	backgroundcolor=\color{verylightgray},
	extendedchars=true,
	basicstyle=\footnotesize\ttfamily,
	showstringspaces=false,
	showspaces=false,
	numbers=left,
	numberstyle=\footnotesize,
	numbersep=9pt,
	tabsize=2,
	breaklines=true,
	showtabs=false,
	captionpos=b
}

\lstset{language=Solidity}
\label{sol:1}
\begin{lstlisting}[frame=shadowbox]
pragma solidity ^0.4.24;
contract forTest {   
    function TestWhile (uint a, uint b) public {
        uint x = 1;
        while (a < b) {
            x += 1;
        }
    }
}
\end{lstlisting}

We test the contract on three different EVM versions---JavaScript implementation js-evm~\cite{jsevm}, Python implementation Py-EVM~\cite{pyevm}, and C++ implementation aleth~\cite{aleth}. When the $TestWhile$ function is called with parameters $a = 1$, $b = 2$, 
js-evm will continually print trace information, together with the increase of CPU usage and terminate the operation correctly; Py-EVM will print trace information and terminate the operation correctly; aleth will not print trace information, but continue to occupy CPU resources until the system kills the whole process and crashes. Comparing the execution results, we find that aleth can neither handle the extreme situation of infinite loop properly that has no friendly user interaction or information feedback nor cut the loss in time and reduce malicious resource occupied, eventually leads to potential denial of service. 

This example shows that some contracts containing corner cases can trigger the boundary condition of EVM implementations and expose unexplored defects. But these contracts involving extreme circumstances are often inconsistent with logic programming rules, which require artificial construction, in other word, contract mutation. Furthermore, for massive mutated contracts, test oracle is difficult to artificially define, and some extreme cases are not designated in EVM design specifications. Therefore, it is an efficient and effective way to apply differential fuzzing.

\begin{figure*}[h]
\centering	
\includegraphics[width=18cm]{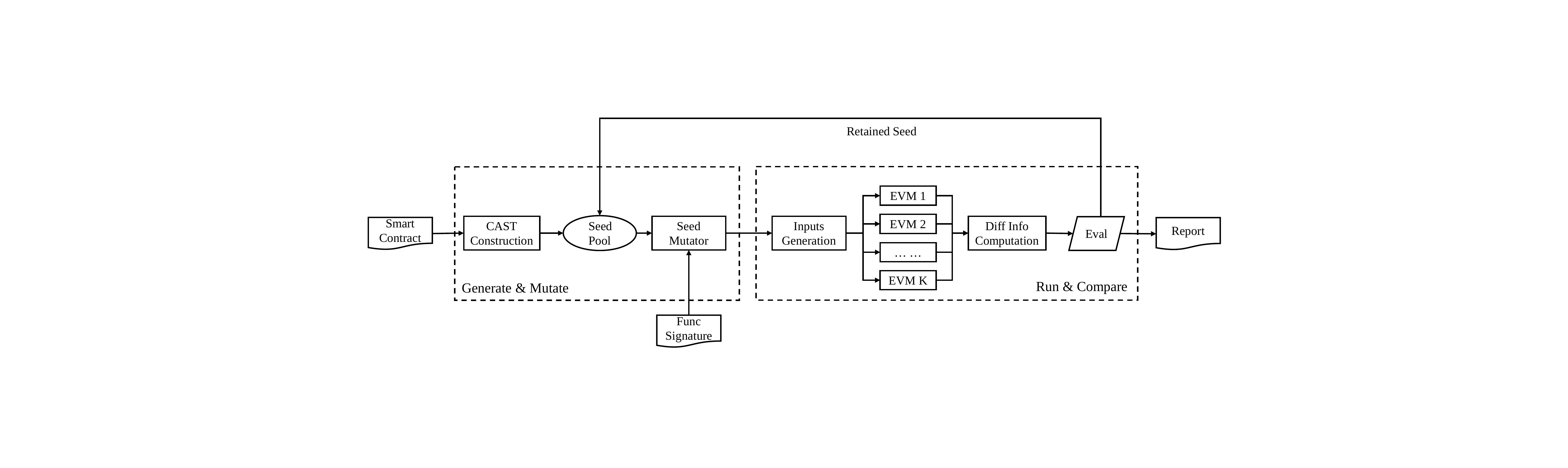}
\vspace{-1cm}
 \caption{The overview of EVMFuzz consists of seed generation module and EVM execution module. The seed module contains seed mutation and selection algorithms, and the EVM module includes multi-EVM running and comparing mechanism.}
\label{overview}
\end{figure*}

\section{Approach Overview}
\label{approach}

In this section, we briefly introduce the workflow of EVMFuzz. 
Our goal is to apply differential fuzz testing on EVMs. The concept of differential fuzz testing is very simple, that is, to continuously provide invalid, unexpected or random data as inputs to several programs with the same functions. These programs are then monitored for catching "different act" on some inputs, if so, we may find a bug in some of the programs. 
In this paper, our test object are the same functional EVM platforms implemented by different programming languages, and the test input is the mutated smart contract. 
An overview of EVMFuzz is given in Fig. \ref{overview}, which consists of two major components, i.e., seed contract generation based on static analysis and unified EVM execution based on fuzzing loop. We will also introduce the evaluation metrics for EVM differential fuzzing.

\subsection{Seed Contract Generation}
The input for the seed generation module is the smart contract file, and the output is a contract variant whose key property has been modified by specific mutators. 
First, we precisely construct the Critical locations identified Abstract Syntax Tree (CAST) of the seed contract (\S \ref{CAST}), for facilitating subsequent mutation and analysis. Then the seed contract will be put into the seed pool. EVMFuzz will rank the candidate contracts as a prioritized queue under the guidance of dynamic priority, and the contract in the first place will be selected as the next subject (\S \ref{SeedContractPrior}). After choosing the contract for mutation, EVMFuzz uses 8 predefined mutators and the combined strategy to guide mutation (\S \ref{ContractMutation}) and obtains the input for unified EVM execution module. The goal is to generate contracts that can increase the degree of metric difference and trigger different execution output.

\subsection{Unified EVM Execution}
EVM execution module provides a unified runtime environment for various EVMs (\S \ref{UnifiedExecution}). After receiving the contract file from the seed generation module, it compiles the seed into EVM bytecode. The input parameter is generated according to the data type of the called function, thus the uniform input for each EVM is obtained. Then EVMFuzz automatically runs all EVMs, calculates the difference information according to the test metric, and compares the execution output results. Finally, according to the seed's ability to enhance the degree of metric difference, EVMFuzz decides whether to put the seed contract into the seed pool where high-quality seeds preserved (\S \ref{SeedSelection}). Besides, when the execution output is inconsistent, this module will also record the potential exception for manual root cause analysis. 

\subsection{Metrics Formulation}
To evaluate the performance of each EVM on the test contract, we define the metric on two general indicators. 
As most EVMs are implemented as a transaction-based state machine, and the change of state depends on the sequence of opcode to be executed, the input parameter and gas limit, hence, we use the internal opcode sequence executed and gas used as the two indicators. 

\begin{enumerate}[leftmargin=*]
\item\textbf{opcode sequence.}
Opcode is short for operation code, which is used to describe the part of machine code that performs some sort of operations in machine language instructions. From the perspective of computer instruction execution, each function call is completed by a series of opcode execution. The opcode sequence clearly shows a complete process of contract operation, which can be used to check the execution correctness of each step. For platform $i$, we define $opSeqLen(i, C)$ as the length of opcode sequence of $i$ when executing contract $C$.

\item \textbf{gasUsed.}
$gasUsed$ is the total number of gas consumed by all operations in a transaction or message. The value of $gasUsed$ is vitally interrelated with the success of transaction execution, and is also directly related to the transaction fee that users ultimately need to pay. Here we use $gasUsed(i, C)$ to represent total gas consuming of platform $i$ after running contract $C$.




\end{enumerate}

Based on these two indicators, we further define the evaluation metric of difference information. When given an input parameter, the normal execution of a transaction on a dedicated EVM platform is determined by a confirmed and unique execution sequence, and the total gas consumption is also calculated. 
Therefore, we construct an evaluation metric $diff$ to measure the difference among different EVMs execution(\S \ref{SeedSelection}). The greater the metric difference, the higher probability the inconsistent execution output. Execution output is the return value after all executions, that $output(i, C)$ is defined as the returns of $C$'s execution on EVM $i$. For a function call, it is the returned data, and for a transaction, it is the balance. While the metric defined on the two internal indicators reflects the implementation and execution difference of different EVMs, the execution output can intuitively reflect whether those EVMs are running consistently or correctly. 

\section{EVMFuzz Design}
\label{Methodology}

In this section, we will elaborate on the key components in Fig. \ref{overview}.

\subsection{CAST Construction}
\label{CAST}
Before EVMFuzz starts the entire procedure of fuzzing, it first carries out static analysis on initial seed contracts and generates the CAST structure for further mutation.

A CAST of a smart contract is a structured tree representation of the abstract syntactic structure of Solidity source code. Each node of the tree denotes a construct occurring in the source code. CAST can define and decompose properties in all statements of contract. Transforming a contract into CAST structure can help us complete the subsequent contract mutation operations. It can directly search, replace, delete or insert operators according to the key attributes. 

\begin{figure}[h]
    \centering
    \includegraphics[width=8cm]{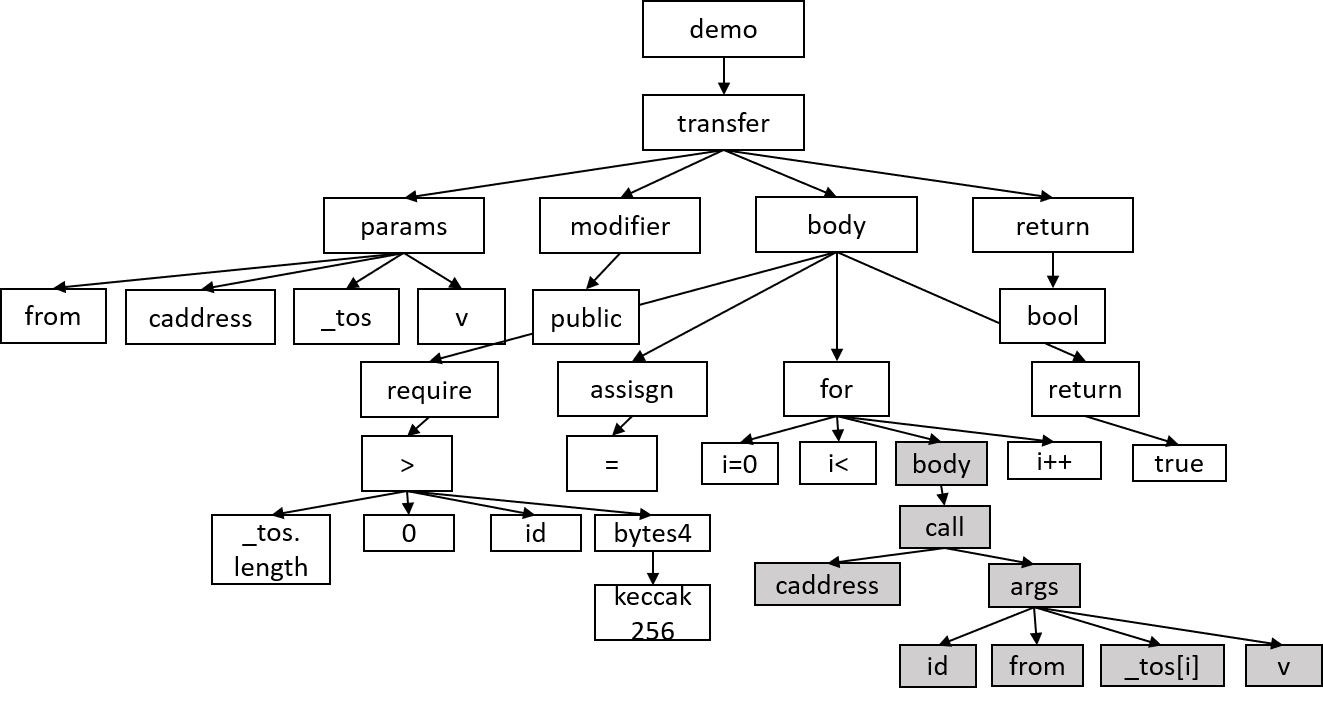}
    \vspace{-0.4cm}
    \caption{CAST structure of Demo contract, and the shaded nodes are the identified critical locations.}
    \label{castExample}
\end{figure}

Furthermore, CAST identifies critical locations of a seed contract, which are the subtree of statements related to ether transaction. It mainly involves six statement symbols --- $new$, $call$, $delegatecall$, $callcode$, $send$ and $transfer$. 
Based on CAST, we can guide pre-defined mutators to select the structures that are identified as critical locations in order to test the core functions of EVM. 
A simple example is presented in Fig.\ref{castExample}, including the source code and the corresponding CAST structure, where the shaded nodes are regarded as the critical locations, for the reason that they are all under the subtree of the $call$ statement.


\lstset{language=Solidity}
\label{sol:2}
\begin{lstlisting}
pragma solidity ^0.4.24;
contract Demo {
    function transfer (address from,address caddress,address[] _tos,uint v) public returns (bool) {
        require(_tos.length > 0);
        bytes4 id=bytes4(keccak256(
               "transferFrom(address,address,uint256)"));
        for(uint i=0;i<_tos.length;i++) {
            caddress.call(id,from,_tos[i],v);
        }
        return true;
    }
}
\end{lstlisting}

\subsection{Seed Contract Prioritization}
\label{SeedContractPrior}
In seed contract pool, the importance of each candidate contract is different. In general, the contract that makes the metric difference among EVMs larger should be the benchmark for the next mutation iteration. But at the same time, in order to ensure the diversity, other contracts should also have a certain probability of being selected. Therefore, we use the dynamic priority scheduling algorithm to maintain a candidate queue. For each contract, we give it an initial priority, and then its value changes with the increasing of waiting time to ensure that every seed has the chance to be selected. 

\begin{algorithm}
  \caption{Seed Prioritization}
  \label{alg:1}
  \KwIn{$candidate\_seeds$ $\leftarrow$ list of candidate contracts\\\quad \quad \quad $diff\_pri$ $\leftarrow$  difference priority of each seed contract \\\quad \quad \quad $time\_pri$ $\leftarrow$ time priority of each seed contract}
  
  \KwOut{$C$ $\leftarrow$ test contract for next iteration}

  $initial$ priority = \{\}\\
  \For{$i$ = $0$ to $len(candidate\_seeds)$}
  {
    cur = candidate\_seeds[i]\\
    priority[cur] = diff\_pri[cur] + time\_pri[cur]
  }
  candidate\_seeds.sort(cmp=priority, reverse=True)\\
  C = candidate\_seeds[0]\\
  /* update time priority */\\
  \For {$i$ = $1$ to $len(candidate\_seeds)$}
  {
    time\_pri[candidate\_seeds[i]] += 1
  }
\end{algorithm}  

\begin{algorithm}
  \caption{Contract Mutation}
  \label{alg:2}
  \KwIn{$C$ $\leftarrow$ contract before mutation\\\quad \quad \quad$diff$ $\leftarrow$ difference information of last time iteration\\\quad \quad \quad$choice$ $\leftarrow$ selection of combined strategy\\\quad \quad \quad$mutator\_list$ $\leftarrow$ list of mutators}

  \KwOut{$C'$ $\leftarrow$ contract after mutation}
  
  C\_tree = generate\_CAST(C)\\
  mutator\_list.sort(cmp=diff)\\
  m = select\_mutator(mutator\_list, choice)\\
  \For {$i$ = $1$ to $len(m)$ }
  {
    C\_tree' = mutate(C\_tree, m[i])
  }
  C' = generate\_code(C\_tree')

\end{algorithm}

As Algo. \ref{alg:1} shows, the priority of each seed contract consists of two parts (Algorithm \ref{alg:1} line 3-4). The first part is metric difference priority, and the initial value is a number between 0 and 10, which is proportional to the value of difference; the second part is time priority, and the initial value is 0. 
Then, all candidate seed contracts are sorted according to the priority value, and the contract with the highest integrated priority is selected as the next mutation object, and the time priority of other seed contracts is increased for next iteration (Algorithm \ref{alg:1} line 9-11). 

\begin{table*}[htp]
\caption{Typical Mutators}
\begin{center}
\begin{tabular}{m{30mm} | m{50mm} | m{85mm}}
  \hline
  \makebox[30mm][c]{\textbf{Object}} & \makebox[50mm][c]{\textbf{Description of Mutator}} & \quad \quad \quad \tabincell{c}{\textbf{Example (contract source code before mutation}\\ \textbf{$\rightarrow$ contract source code after mutation)}} \\ \hline
  \makebox[30mm][c]{local variable} & Select a method and modify one or more local variable attribute(s) & \tabincell{l}{function multiPath (uint a, uint b) public returns (uint)\\ \{ \hl{uint} x; \}\\ $\rightarrow$ function multiPath (uint a, uint b) public returns (uint)\\ \{ \hl{bytes} x; \}}\\ \hline
  \makebox[30mm][c]{function property} & Select a method and insert, modify or delete its function attribute(s) & \tabincell{l}{function multiPath (uint a, uint b) \hl{public} returns (uint) \{...\}\\ $\rightarrow$ function multiPath (uint a, uint b) \hl{constant} returns (uint) \{...\}} \\ \hline
  \makebox[30mm][c]{arithmetic operator} & Select a method and modify one or more inner arithmetic operator(s) & \tabincell{l}{function multiPath (uint a, uint b) public returns (uint)\\ \{ uint c = \hl{a + b}; \}\\ $\rightarrow$ function multiPath (uint a, uint b) public returns (uint)\\ \{ uint c = \hl{a - b}; \} } \\ \hline
  \makebox[30mm][c]{conditional operator} & Select a method and modify one or more inner conditional operator(s) & \tabincell{l}{... if (\hl{a > b}) \{ ... \} ... \\ $\rightarrow$ ... if (\hl{a <= b}) \{ ... \} ...} \\ \hline
  \makebox[30mm][c]{loop operator} & Select a method and modify one or more inner loop execution statement(s) & \tabincell{l}{... for (uint i = 0; \hl{i < 100}; ++i) \{ ... \} \\ $\rightarrow$ ... for (uint i = 0; \hl{i < 199}; ++i) \{ ... \}} \\ \hline
  \makebox[30mm][c]{assert statement} & Select a method and insert or delete one or more assert statement(s) & \tabincell{l}{function sub (uint a, uint b) public returns (uint)\\ \{ ... \}\\ $\rightarrow$ function sub (uint a, uint b) public returns (uint)\\ \{ ... \hl{assert (a > b)}; ... \} } \\ \hline
  \makebox[30mm][c]{return statement} & Select a method and delete the return statement & \tabincell{l}{function get (uint a) public \hl{returns (uint)}\\ \{... \hl{return a;} ...\}\\ $\rightarrow$ function get (uint a) public\\ \{ ... \} } \\ \hline
  \makebox[30mm][c]{control structure} & Select a method and insert \textbf{continue} or \textbf{break} statement in loop structure & \tabincell{l}{... for (uint i = 0; i < 10; ++i) \{ a += b; \}\\ $\rightarrow$ for (uint i = 0; i < 10; ++i) \{ a += b; \hl{break;} \}} \\ \hline
\end{tabular}
\label{mutators}
\end{center}
\end{table*}

\subsection{Seed Contract Mutation}
\label{ContractMutation}
The goal of EVMFuzz is to generate high-quality seed contracts that can trigger more discrepancies among EVMs and to identify vulnerabilities in EVMs. 
As Algo. \ref{alg:2} shows, EVMFuzz first 
generates CAST from contract's Solidity code (Algorithm \ref{alg:2} line 1). Then, it updates the weight of each mutator based on the metric difference feedback (Algorithm \ref{alg:2} line 2), and selects a combination strategy to mutate the candidate seed contract with the corresponding mutators (Algorithm \ref{alg:2} line 3-6). Finally, soltar~\cite{soltar} is utilized to reconstruct the code from CAST (Algorithm \ref{alg:2} line 7). 
Details about the mutators and mutation strategies are introduced as below.

\noindent \textbf{Typical Mutators.} \quad
During mutation, we must ensure the syntax correctness of the modified contract so that it can generate bytecode normally. Currently, we design 8 mutators according to the functional logic features of the smart contract, shown in Table \ref{mutators}. These variations are based on three different granularity, the first is word-level (row 2-3), modifying variable and function attributes, which could influence the storage and syntax structure; the second is character-level (row 4-6), modifying the arithmetic or conditional operators and the terminate condition in a loop, which may stochastically change the control flow; and the last is statement-level (row 7-9), like insert or delete some assert statements which are used for internal condition judgment, etc.

\noindent \textbf{Mutator Selection.} \quad
Each mutator performs differently, so we can also maintain a priority queue based on the feedback metric difference. For a seed contract, we update the weight of corresponding mutators after the multi-version EMVs comparison, which is similar to the initialization of metric difference priority (\S 4.2). If the metric difference increases, the mutator ID is pushed into the queue in descending order according to the weight; otherwise, the queue will not update. 
Except for the weight update, we design five mutator combined strategies to further increase the randomness and diversity of mutation in each iteration:

\begin{itemize}[leftmargin=*]
\item OddComb: Combination of the mutators whose index is odd. 
\item EvenComb: Combination of the mutators whose index is even. 
\item ExtremeComb: Combination of first and last mutator. 
\item RandomComb: Randomly select a mutator, without weight. 
\item AllComb: Randomly choose one strategy above in each iteration. 
\end{itemize}

\subsection{Unified EVM Execution}
\label{UnifiedExecution}

After an effective contract mutation, it is necessary to compare the metric difference of the EVMs' execution in order to guide the subsequent mutation and selection of seed contracts. The execution of different EVMs requires unified management. 

The first step is to get the data that can directly feed into EVM platform, namely the contract bytecode and input parameters. By executing $solc\ -bin\ -runtime\ xx.sol$, we can compile the mutated seed contract $C$ into runtime bytecode. According to the selected function's received data type, we can generate the input parameters. For each data type, we pre-define some common or extreme values, which are randomly selected at the time of generation. 
The second step is to call the execution interface of each EVM to execute the contract data, standardize the output format of the execution result and save the output. 
The third step is to analyze the output files of each platform, comparing the data of each indicator and calculating the metric difference for next iteration.  

\subsection{Seed Contract Selection}
\label{SeedSelection}
As mentioned above, we evaluate the quality of a generated seed contract based on platform diversity, so we select candidate seed $C$ based on the metric difference $diff$, which is defined as follows.
\begin{align}
\label{diff_formula}
&gasDiff(i, j) = norm(abs(gasUsed(i, C),gasUsed(j, C)))\nonumber\\
&opDiff(i, j) = norm(abs(opSeqLen(i, C),opSeqLen(j, C)))\nonumber\\
&diff = gasDiff(i, j) + opDiff(i, j) \nonumber\\
&outVul(i, j) = output(i, C) \oplus output(j, C)\nonumber
\end{align}

Difference is used to measure the input seed's ability of inducing platforms to make differential decisions. If a mutated contract enlarges the difference after executing on different EVMs, we consider it to be a high-quality seed with a higher probability of triggering vulnerabilities and store it in the seed queue as a candidate for the next iteration of prioritization and mutation (Algorithm \ref{alg:4} line 2-7). 

\begin{algorithm}
  \caption{Seed Selection}
  \label{alg:4}
  \KwIn{$C'$ $\leftarrow$ contract after mutation\\\quad \quad \quad $record$ $\leftarrow$ maximum difference information recorded}

  \KwOut{$candidate\_seeds$ $\leftarrow$ list of candidate contracts\\\quad \quad \quad \,\,\, $diff\_pri$ $\leftarrow$ difference priority of each seed contract\\\quad \quad \quad \,\,\, $time\_pri$ $\leftarrow$ time priority of each seed contract}
  
  diff = run(C')\\
  \If {diff > record}
  {
    diff\_pri[C'] = 10 * normalize(diff)\\
    time\_pri[C'] = 0\\
    candidate\_seeds.append(C')\\
    record = diff
  }

\end{algorithm}

\section{Evaluation}
\label{Evaluation}

In this section, we present the experiment details. EVMFuzz has conducted large-scale mutations on 36,295 real-world smart contracts and several EVM discrepancies and vulnerabilities have been found. We answer the following three questions: \textbf{(i) Is there any inconsistency among EVMs?  (ii) Could EVMFuzz generate high-quality seed contracts efficiently? (iii) Is it possible to find EVM bugs through differential fuzz  testing?}

\subsection{Data and Environment Setup}
All experiments were performed atop a machine with 8 cores (Intel i7-7700HQ @3.6GHz), 16GB of memory, and Ubuntu 16.04.4 as the host operating system. The 36,295 real-world contracts we used for mutation were crawled from the Etherscan~\cite{etherscan} and we obtained the corresponding bytecode and abi function via the Solidity language compiler solc 0.4.24~\cite{Solidity}. EVMFuzz tested four widely used EVM platforms for each mutated contract, that is, js-evm~\cite{jsevm}, Py-EVM~\cite{pyevm}, aleth~\cite{aleth} and geth~\cite{geth}. Considering the fairness, all platforms were running in the same environment. 

All those initial real-world contracts are deployed in the Ethereum and have different addresses. We first analyzed them in four aspects: lines of code, functions, all relevant opcodes and some specific opcodes.
More than 6,500 contracts have over 500 lines of code, and 90 of them have more than 2,000 lines of code, which shows that the target contracts we choose are complicated. Since our experiment is based on the function level, every specific function in the contract is of great significance to our work. 
The total function number runs up to 1,013,013 and 83\% contracts have more than 15 inner functions, which provides a wide range of space for mutation.

In essence, EVM executes opcode sequence rather than the source code written in high-level languages, whose completeness is closely related to the ability to handle distinct opcodes. 
The number of instructions in a single contract ranges from 32 to 83,329, with an average of 6,472. In order to ensure the normal execution of all transactions, each EVM must correctly handle the logic of 140 different instructions. 
The statistic on the occurrences of 11 representative opcodes in each contract is analyzed. The opcodes that related to stack (like PUSH1, POP, DUP1) or conditional jumps (like AND, ISZERO, JUMPDEST) have a higher frequency of occurrence, which has also been considered in the mutators design. 

\subsection{Is there any inconsistency among EVMs?}
From the most intuitive statistics, among those real-world contracts, 34,699 contracts were successfully executed by four platforms and the execution outputs were the same. 
Based on these initial contracts, we will elaborate on the metric difference among different EVMs in the view of the two internal test indicators: $gasUsed$ and opcode sequence. 

\begin{figure*} 
  \centering 
  \subfigure[js-evm]{ 
    \label{dataset:a}
    \includegraphics[width=4cm]{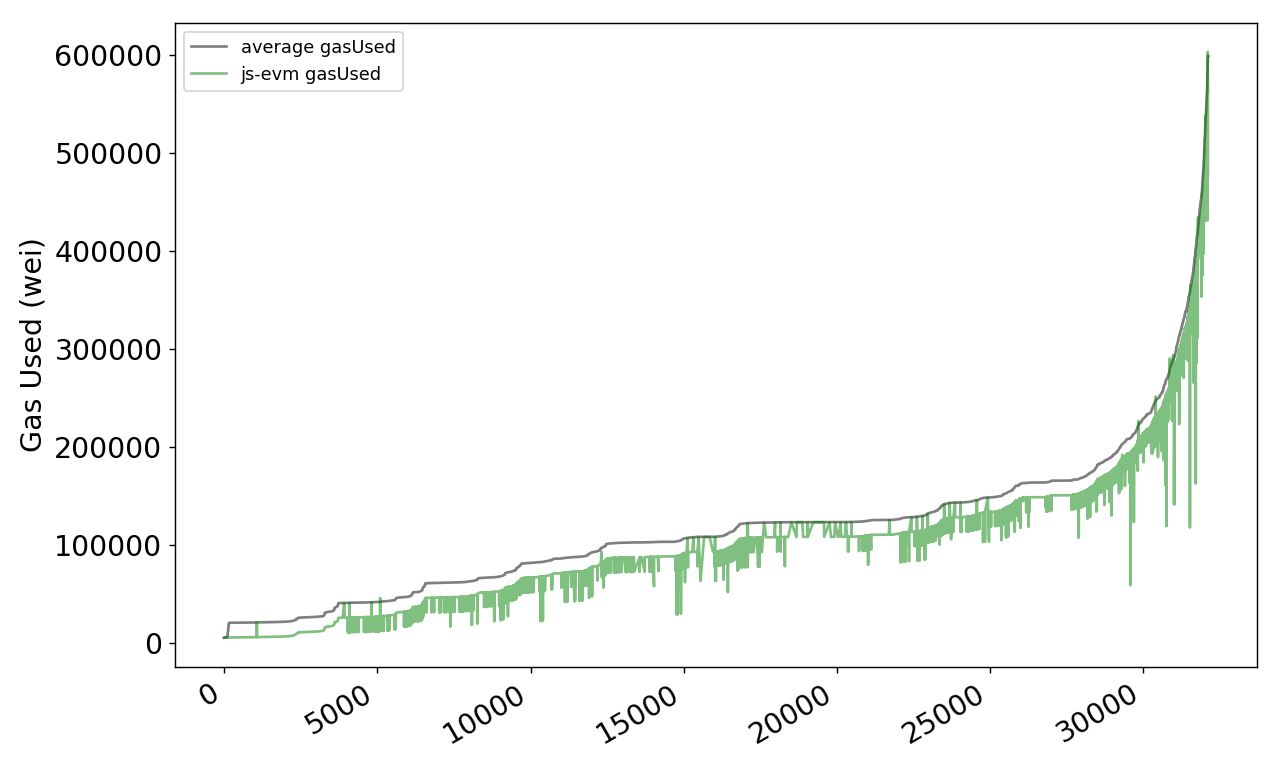}
  } 
  \subfigure[Py-EVM]{ 
    \label{dataset:b}
    \includegraphics[width=4cm]{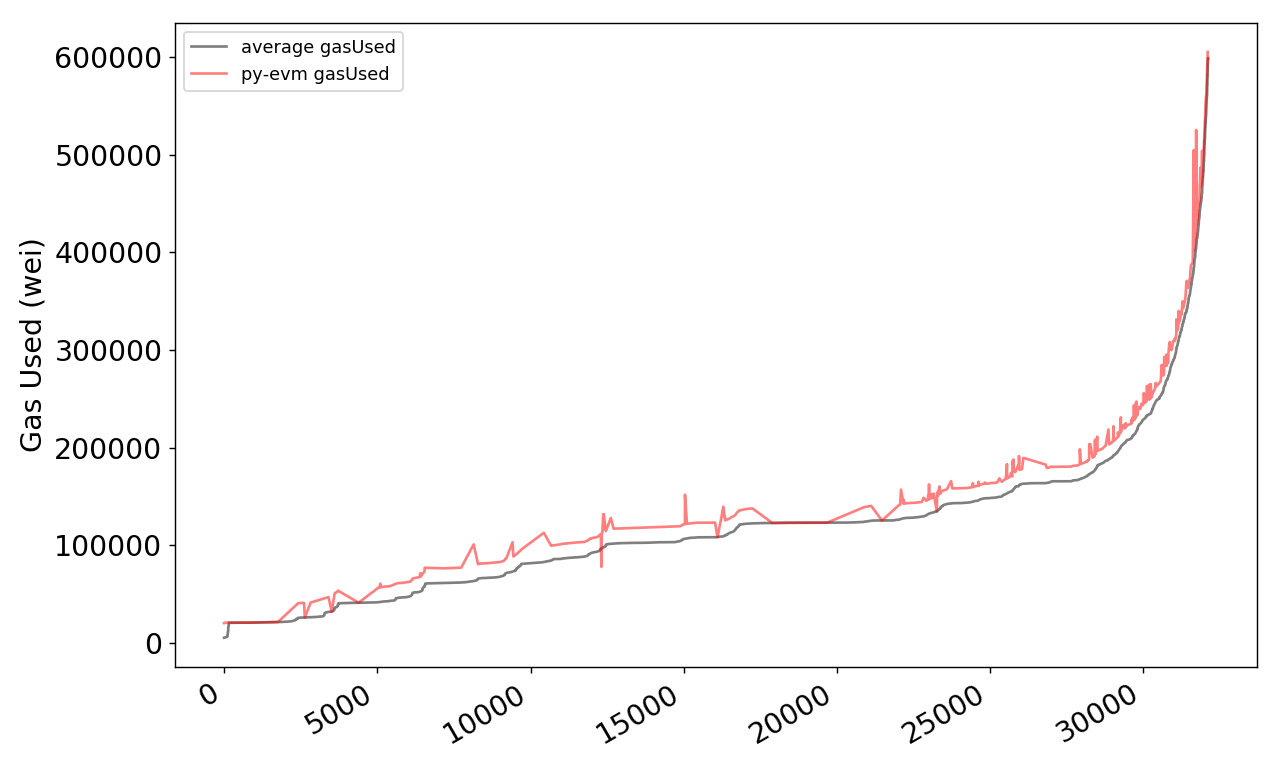}
  } 
  \subfigure[aleth]{ 
    \label{dataset:c}
    \includegraphics[width=4cm]{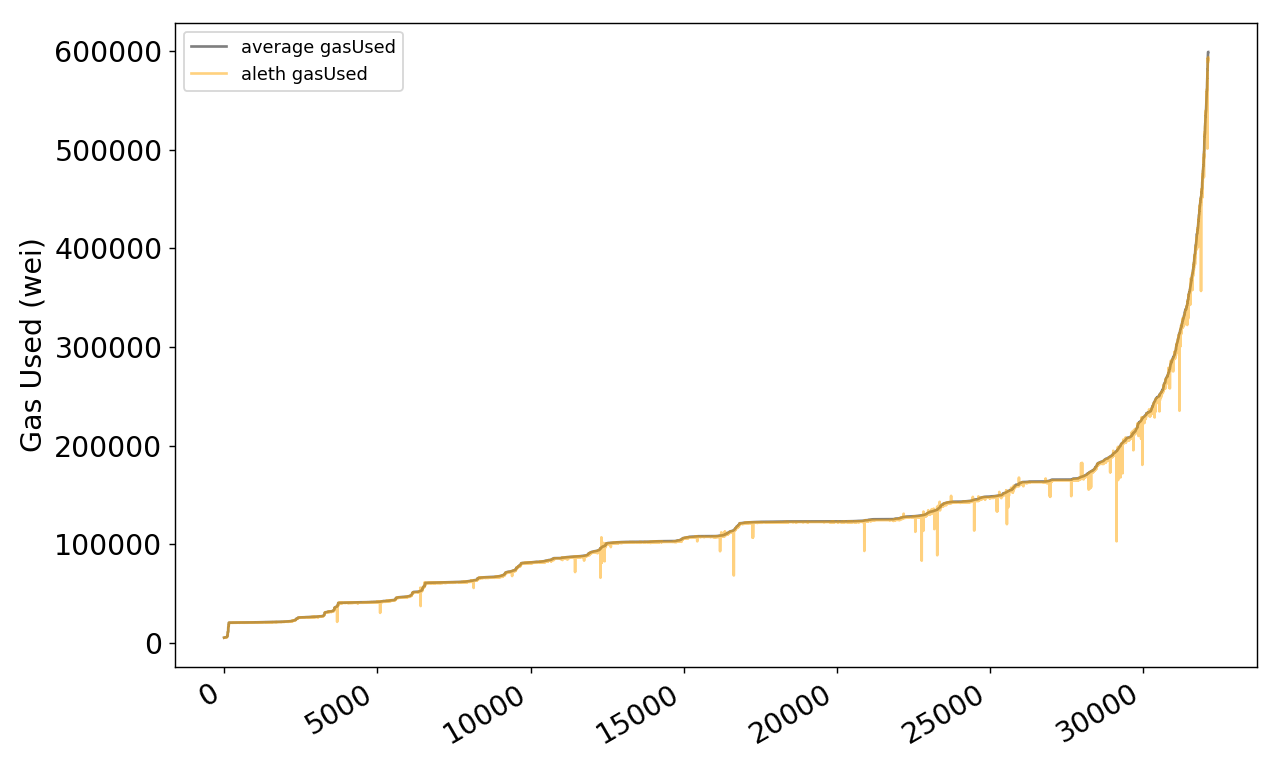}
  } 
  \subfigure[geth]{ 
    \label{dataset:d}
    \includegraphics[width=4cm]{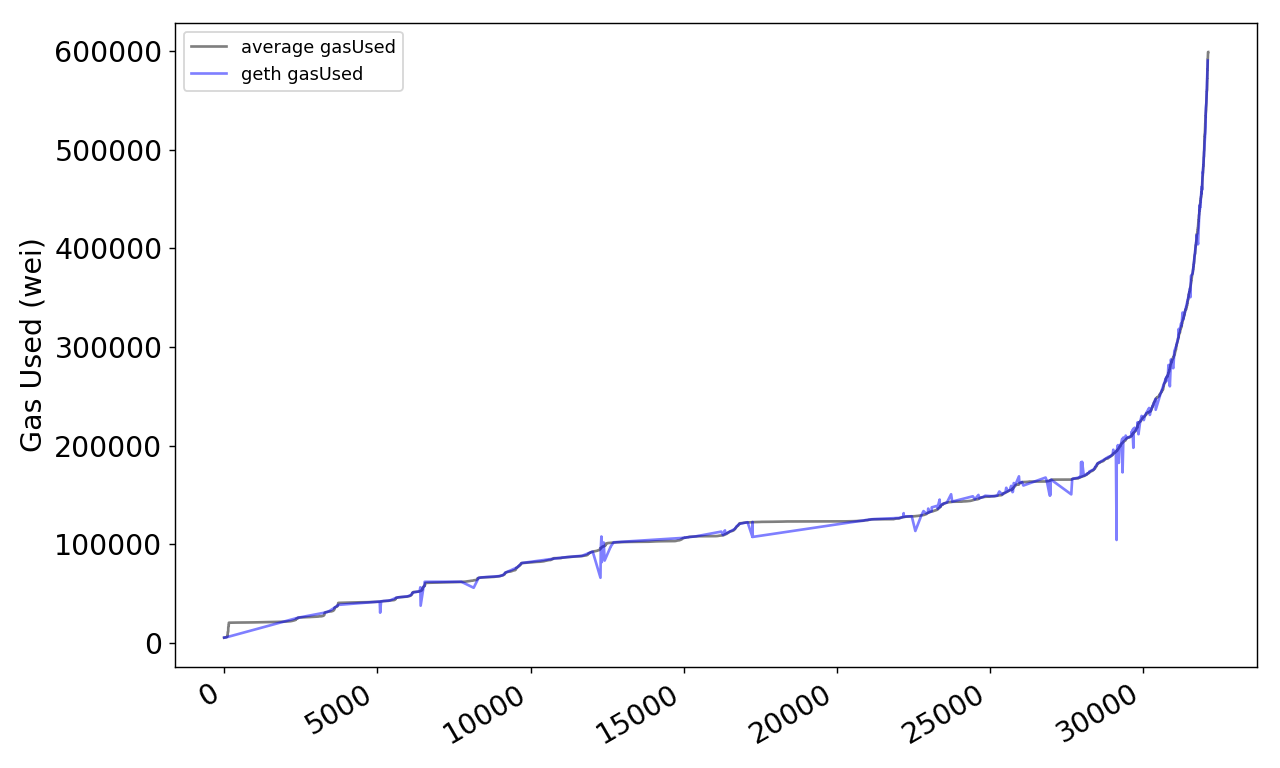}
  } 
    \vspace{-0.5cm}
  \caption{GasUsed inconsistency of different EVM platforms.} 
  \label{gasUsed}
\end{figure*}

\subsubsection{gasUsed inconsistency}
~\\
\indent Within the 34,699 successfully executed contracts, there are 33,424 contracts, which are executed normally with the same opcode sequence. They are used for $gasUsed$ comparison, which excludes the inconsistency of gas consumption caused by different execution opcode sequences. 
Table \ref{gas-diff} is a statistic of the number of contracts with inconsistencies among different EVMs. 

\begin{table}[htp]
\caption{gasUsed inconsistency}
\vspace{-0.5cm}
\begin{center}
\begin{tabular}{ | c | c | c | c | c | c | c | }
  \hline
   & js-evm & Py-EVM & aleth & geth & total & percent \\ \hline
  js-evm & 0 & 18115 & 31166 & 17486 & 66767 & 66.59\% \\ \hline
  Py-EVM & 18115 & 0 & 31176 & 1358 & 50649 & 50.51\% \\ \hline
  aleth & 31166 & 31176 & 0 & 31163 & 93505 & 93.25\% \\ \hline
  geth & 17486 & 1358 & 31163 & 0 & 50007 & 49.87\% \\ \hline
\end{tabular}
\label{gas-diff}
\end{center}
\end{table}

In the process of contract execution, almost every platform has over 50\% average inconsistency rate of $gasUsed$ with others, and aleth even produces a different gas consumption over 90\% of contracts. There are two reasons for this phenomenon. For one reason, the $gasCost$ for some same opcodes defined by different EVM platforms is inconsistent. But for the reason that all EVMs are implemented based on the Yellow paper, this situation rarely occurs. For another, the yellow paper did not specify specific gas consumption values for some opcodes, such as MSTORE, SLOAD, JUMPDEST. Then, different platforms adopt different refund mechanisms, which also results in the inconsistency. 

For further analysis, we calculated the average value of these 33,424 smart contracts as the baseline, then the gas consumption of each EVM is compared with it, as presented in Fig. \ref{gasUsed}(a)-(d).
From these figures, js-evm consumed less gas than baseline, while Py-EVM consumed more, and geth and aleth consumed almost the same as average. Based on these observations, developers should pay attention and set different $gasLimit$ accordingly when deploying the same contract on different EVMs to ensure the correct transaction.

\subsubsection{opcode sequence inconsistency}
~\\
\indent When analyzing the experimental results, we found that even the execution outputs of the four EVMs were the same, their opcode sequences are not completely consistent. In total, 1,275 seed contracts were successfully executed on four EVM platforms and return the same output, but the sequence lengths were different. The inconsistency in sequence length may be due to the fact that each EVM platform internally optimizes when executing the bytecode.



\begin{figure*} 
  \centering 
  \subfigure[js-evm]{ 
    \label{dataset:a}
    \includegraphics[width=4cm]{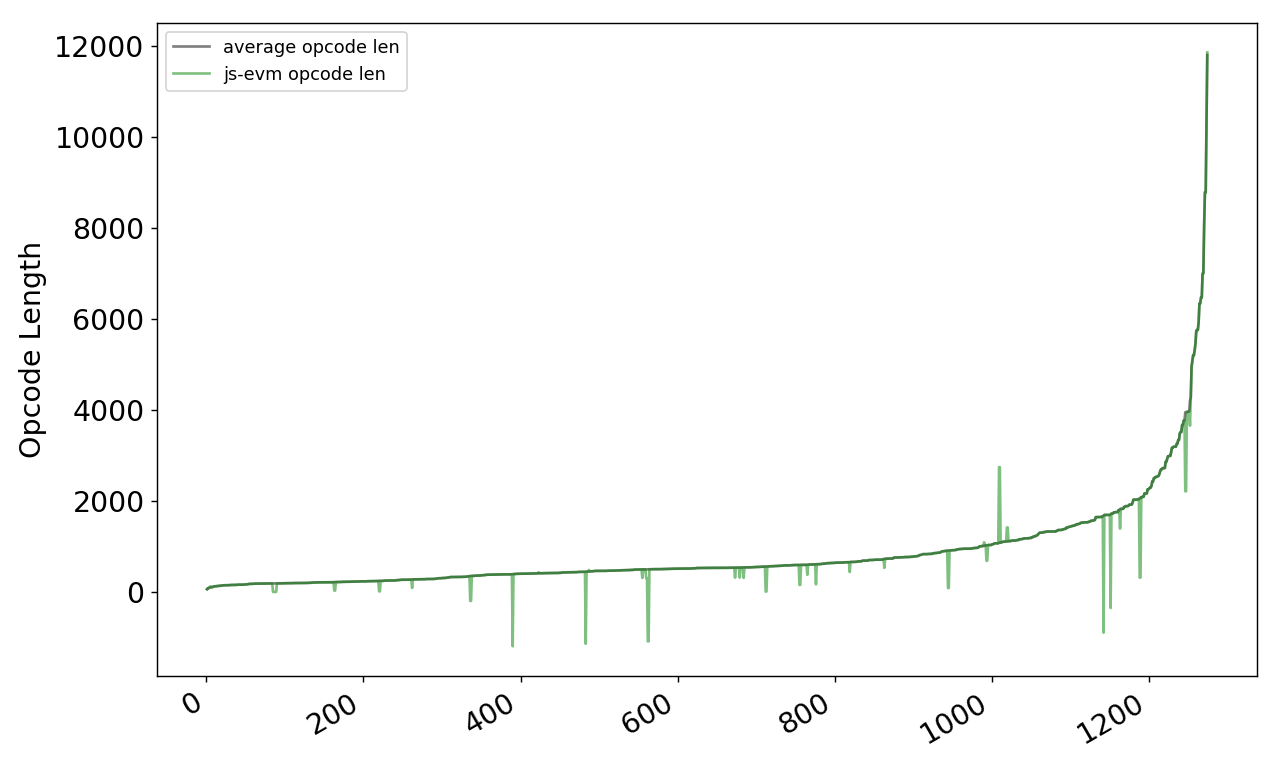}
  } 
  \subfigure[Py-EVM]{ 
    \label{dataset:b}
    \includegraphics[width=4cm]{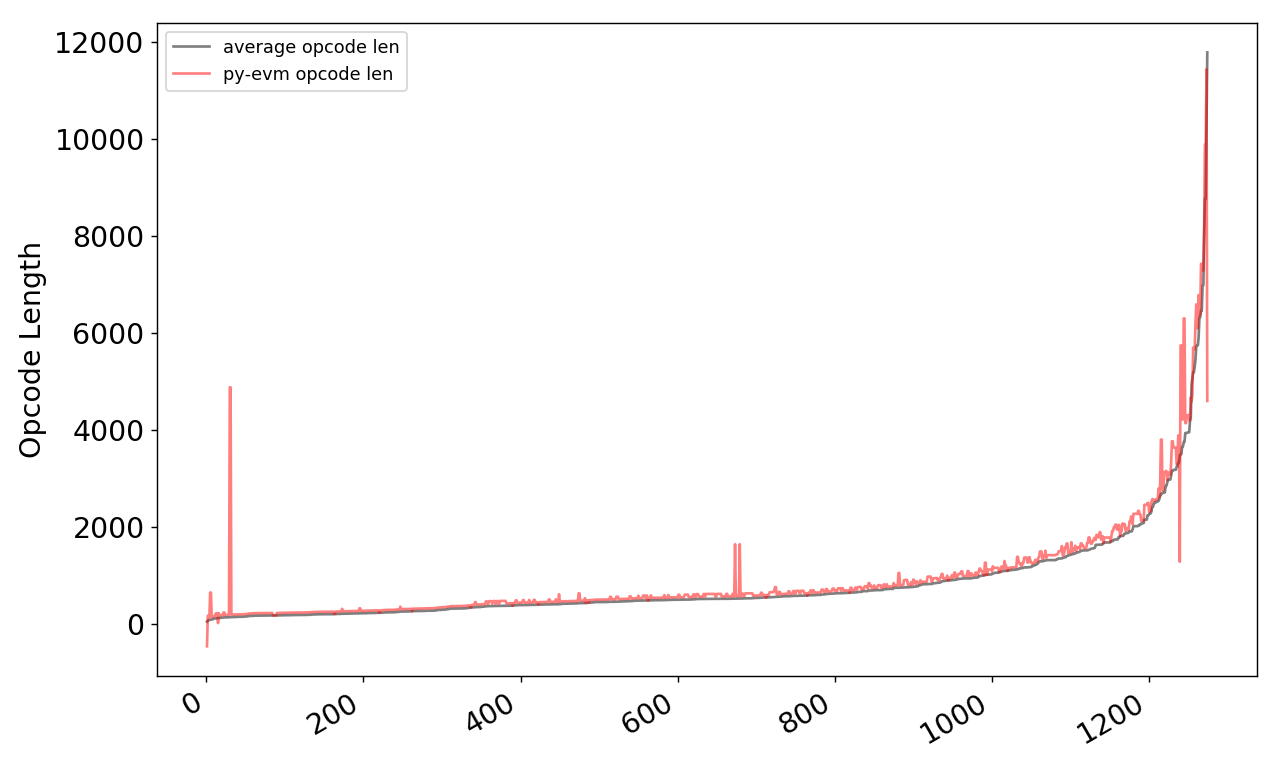}
  } 
  \subfigure[aleth]{ 
    \label{dataset:c}
    \includegraphics[width=4cm]{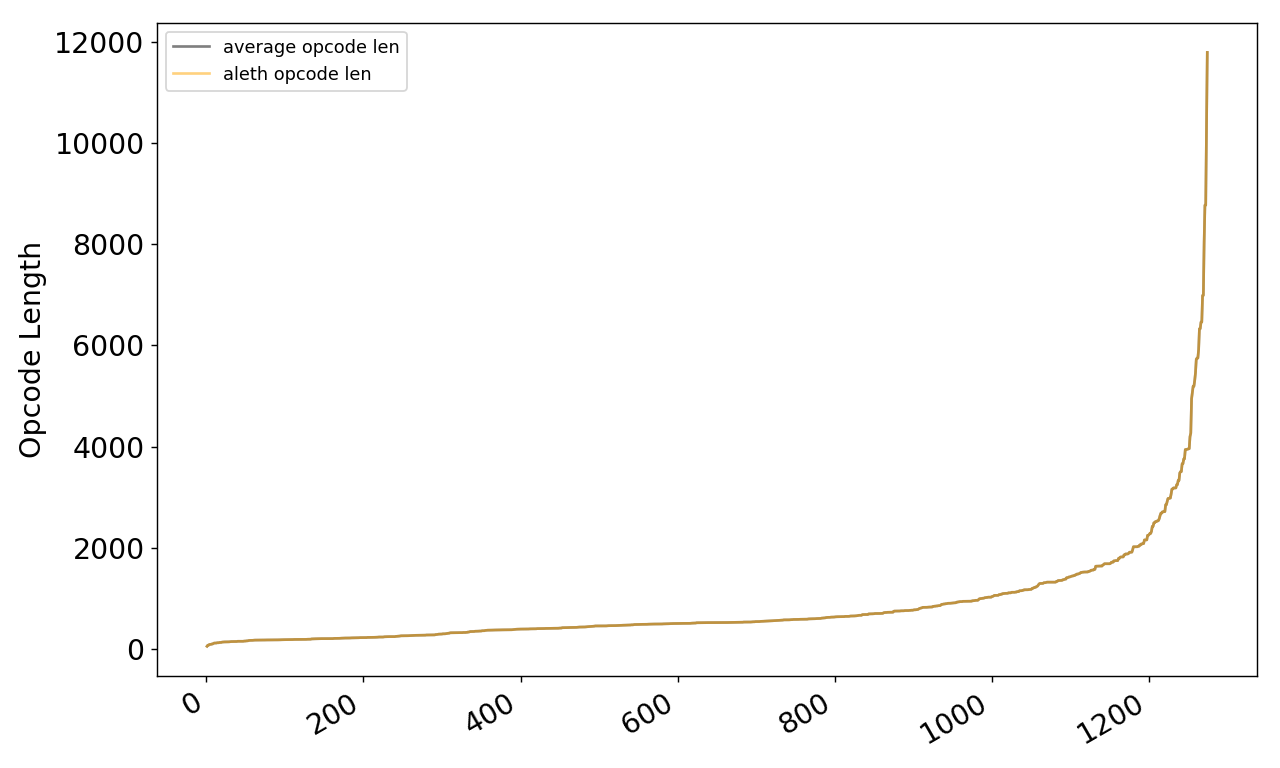}
  } 
  \subfigure[geth]{ 
    \label{dataset:d}
    \includegraphics[width=4cm]{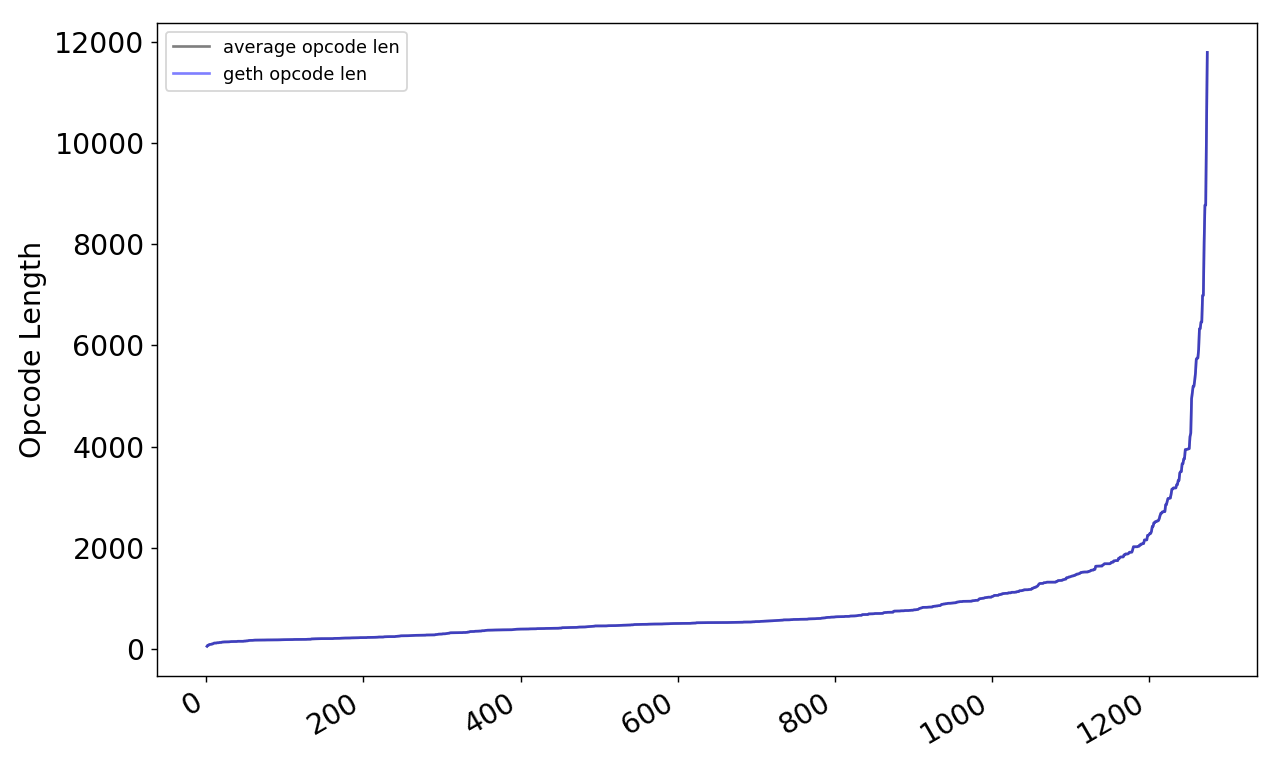}
  } 
    \vspace{-0.5cm}
  \caption{Opcode sequence inconsistency of different EVM platforms.} 
  \label{opcode-fig}
\end{figure*}

\begin{table}[htp]
\caption{opcode inconsistency}
\vspace{-0.5cm}
\begin{center}
\begin{tabular}{ | c | c | c | c | c | }
  \hline
   & js-evm & Py-EVM & aleth & geth  \\ \hline
  js-evm & 0 & 1275 & 52 & 52  \\ \hline
  Py-EVM & 1275 & 0 & 1240 & 1240 \\ \hline
  aleth & 52 & 1240 & 0 & 0 \\ \hline
  geth & 52 & 1240 & 0 & 0  \\ \hline
\end{tabular}
\label{opcode-diff}
\end{center}
\end{table}


From Fig. \ref{opcode-fig}, we can see that the sequence length of geth and aleth on these 1,275 contracts were always the same; the length of the opcode sequence of js-evm is small and always below the baseline; but the length of the execution sequence of Py-EVM is large and always above the baseline. 
From the perspective of execution sequence length, the optimization function of js-evm is better, which reduces the execution burden of EVM to some extent. Those optimizations might be integrated into other EVM implementations. However, considering the following output analysis results, it is possible that this optimization also leads to a reduction in the execution robustness. 

From the above intuitive statistics, it is reasonable to conclude that there are inconsistencies among the implementation and execution of different EVMs, and it is possible to leverage the metric difference of $gasUsed$ and opcode sequence indicator to guide the generation of contracts resulting in potential inconsistent execution output.  

\subsection{Could EVMFuzz generate high-quality seed contracts efficiently?}
Within three days, based on the crawled real-world initial contracts, EVMFuzz generated and executed 253,153 non-redundant contracts. 
Among them, more than half of the seed contracts (66.2\%) successfully trigger the differential performance of $gasUsed$ and opcode sequence indicator, including 1,596 variant contracts showed inconsistent output results among the four EVM platforms. In the process of mutation, the generated seed contracts may contain some rare situations, such as illegal data type or infinite loop, due to the change of key attribute values. If an EVM platform cannot deal with these boundary conditions properly, it will not get the correct output result, and even lead to some serious security problems, as the motivating example presented in the background.

For those 1,596 inconsistencies of the execution output, we refine two metrics for each EVM --- \textbf{Ind.1} is that only this platform does not crash and all others crash; \textbf{Ind.2} is that only this platform crashes and all others do not crash. Through these two numbers, we can see each EVM's ability of keeping good consistency with others and its robustness. These two numbers indicate inconsistencies and there are definitely some potential defects in either the crashed platform or the non-crashed platform. Then, we can further manually analyze the root cause of these potential defects. 
%

\vspace{-0.15cm}
\begin{table}[!htp]
\caption{output inconsistency refinement}
\vspace{-0.5cm}
\begin{center}
\begin{tabular}{ | c | c | c | c | c | }
  \hline
  \makebox[25mm][c]{} & \makebox[10mm][c]{js-evm} & \makebox[10mm][c]{Py-EVM} & \makebox[10mm][c]{aleth} & \makebox[10mm][c]{geth} \\ \hline
   value of Ind.1 & 131 & 0 & 38 & 0 \\ \hline
   value of Ind.2 & 9 & 14 & 0 & 0 \\ \hline
   total inconsistency & 140 & 14 & 38 & 0 \\ \hline
\end{tabular}
\label{Output}
\end{center}
\end{table}

Furthermore, to evaluate the efficiency of different mutation strategies and the quality of each mutator, we selected a contract and its inner function with the longest lines of code from the crawled real-world contracts for detail demonstration. This function receives two parameters and returns a $uint$ type, including two loops and six branch structures. Each branch contains a different number of conditional judgments, arithmetic operations and assert statements, which provides a large mutation space.

\begin{figure*} 
  \centering 
  \subfigure[Mutator \#1]{ 
    \label{diff:a}
    \includegraphics[width=4cm]{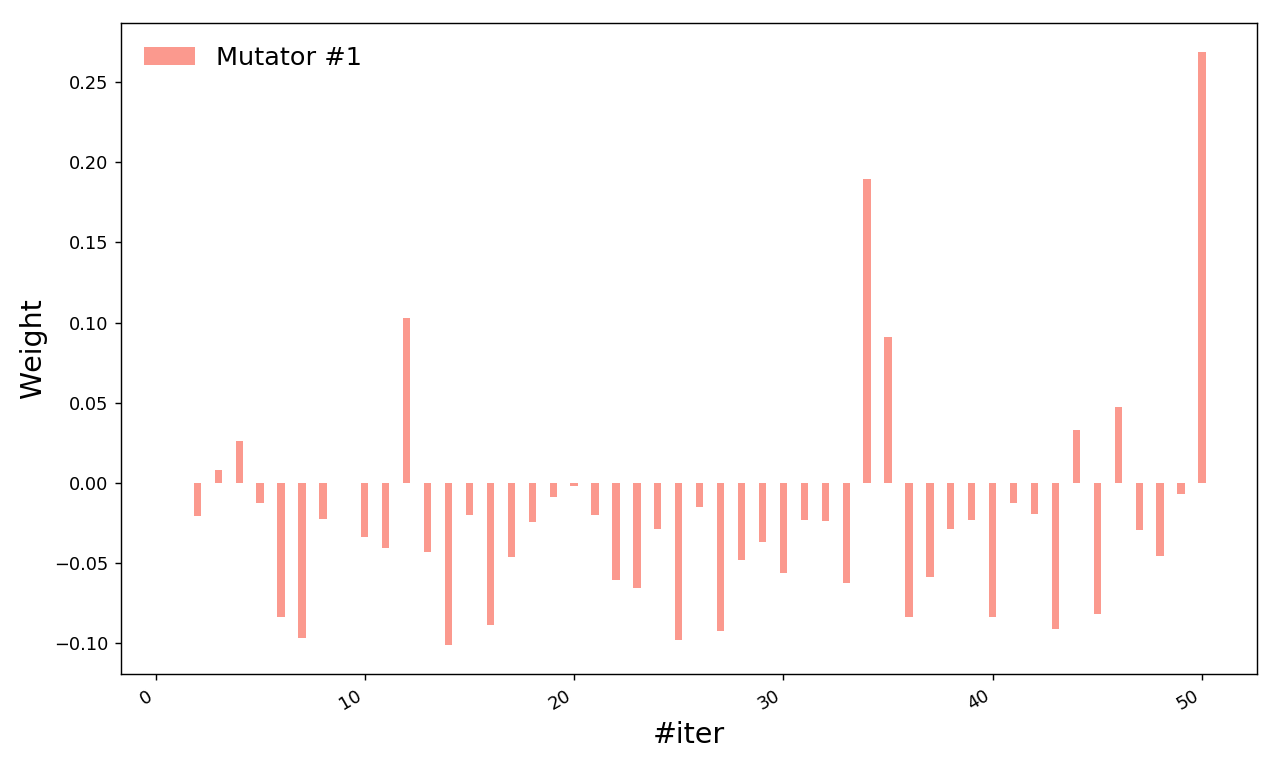}
  } 
  \subfigure[Mutator \#2]{ 
    \label{diff:a}
    \includegraphics[width=4cm]{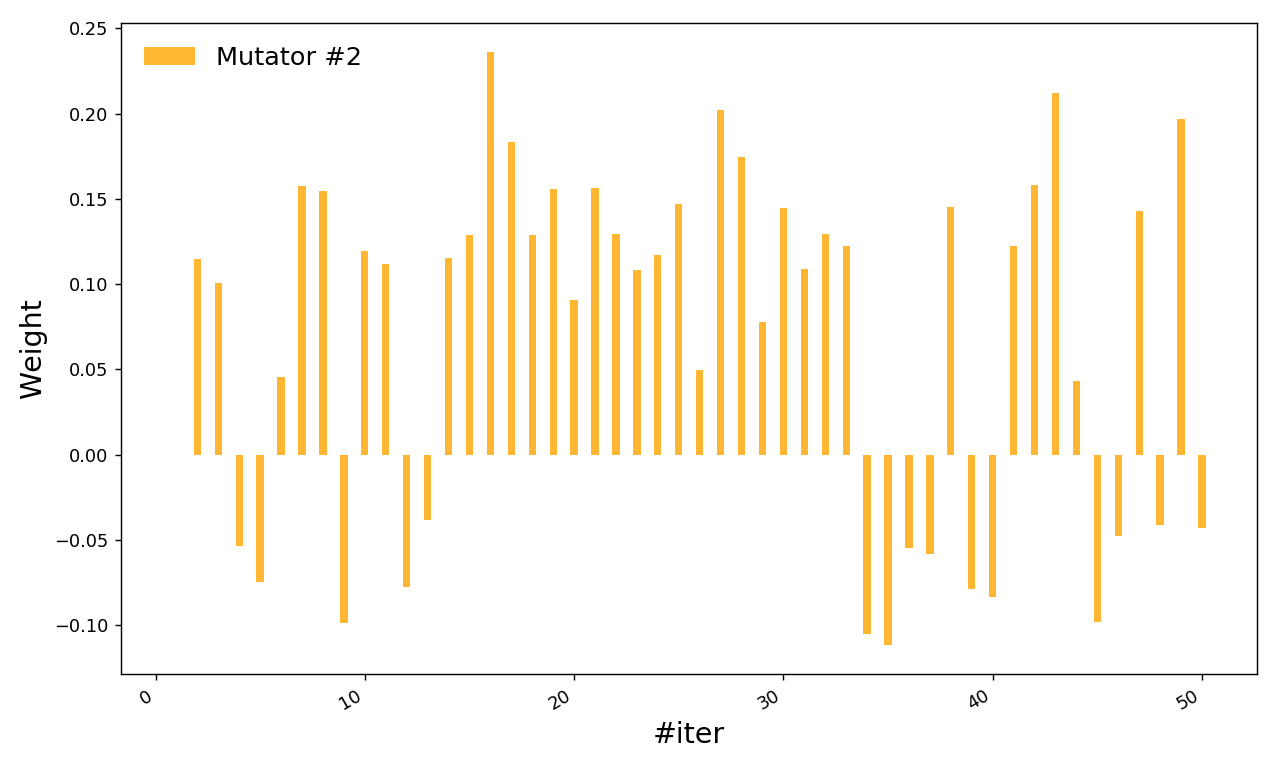}
  } 
  \subfigure[Mutator \#3]{ 
    \label{diff:c}
    \includegraphics[width=4cm]{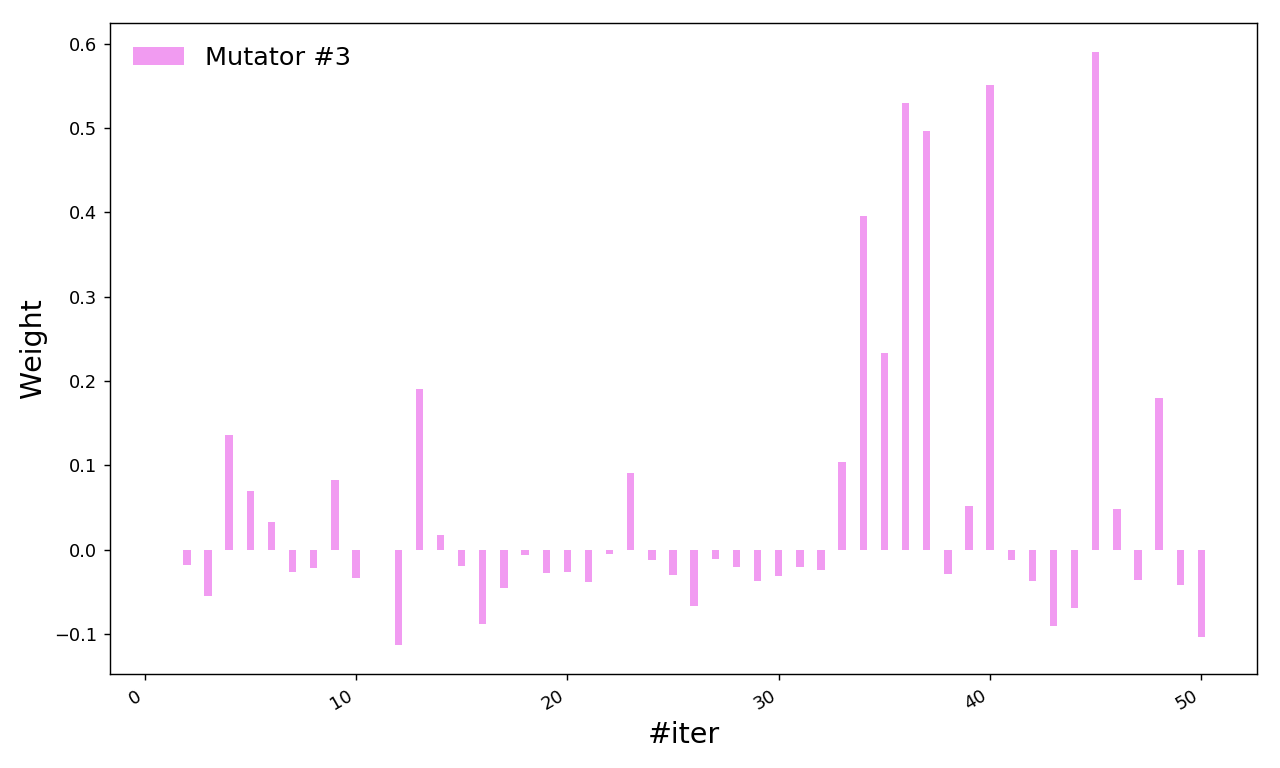}
  } 
  \subfigure[Mutator \#4]{ 
    \label{diff:d}
    \includegraphics[width=4cm]{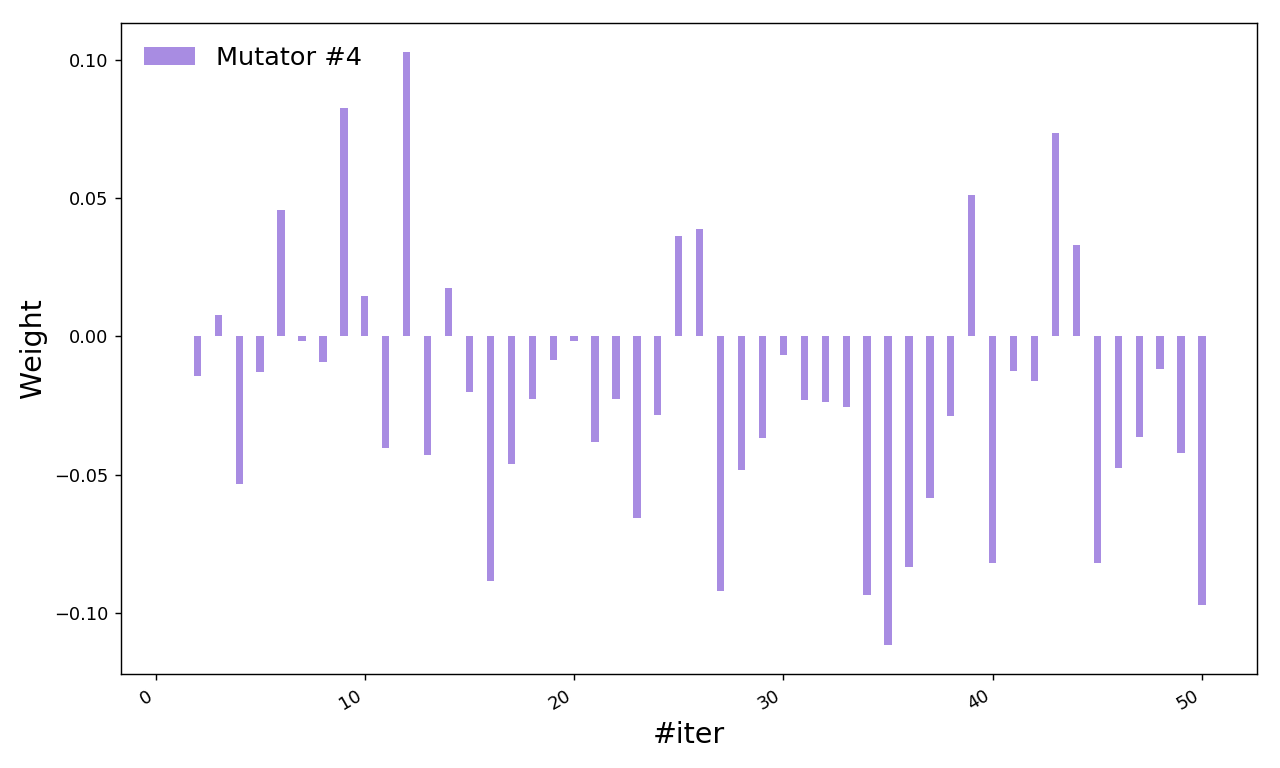}
  } 
  \subfigure[Mutator \#5]{ 
    \label{diff:e}
    \includegraphics[width=4cm]{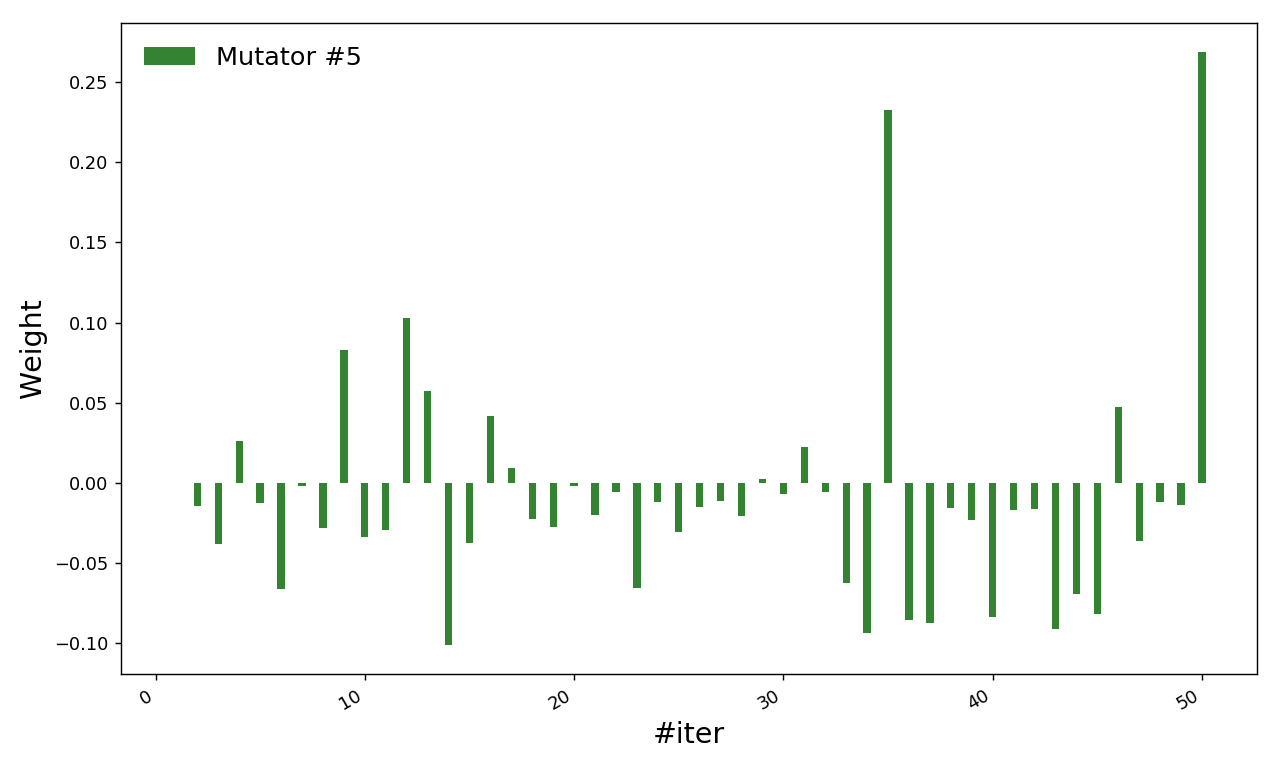}
  } 
  \subfigure[Mutator \#6]{ 
    \label{diff:f}
    \includegraphics[width=4cm]{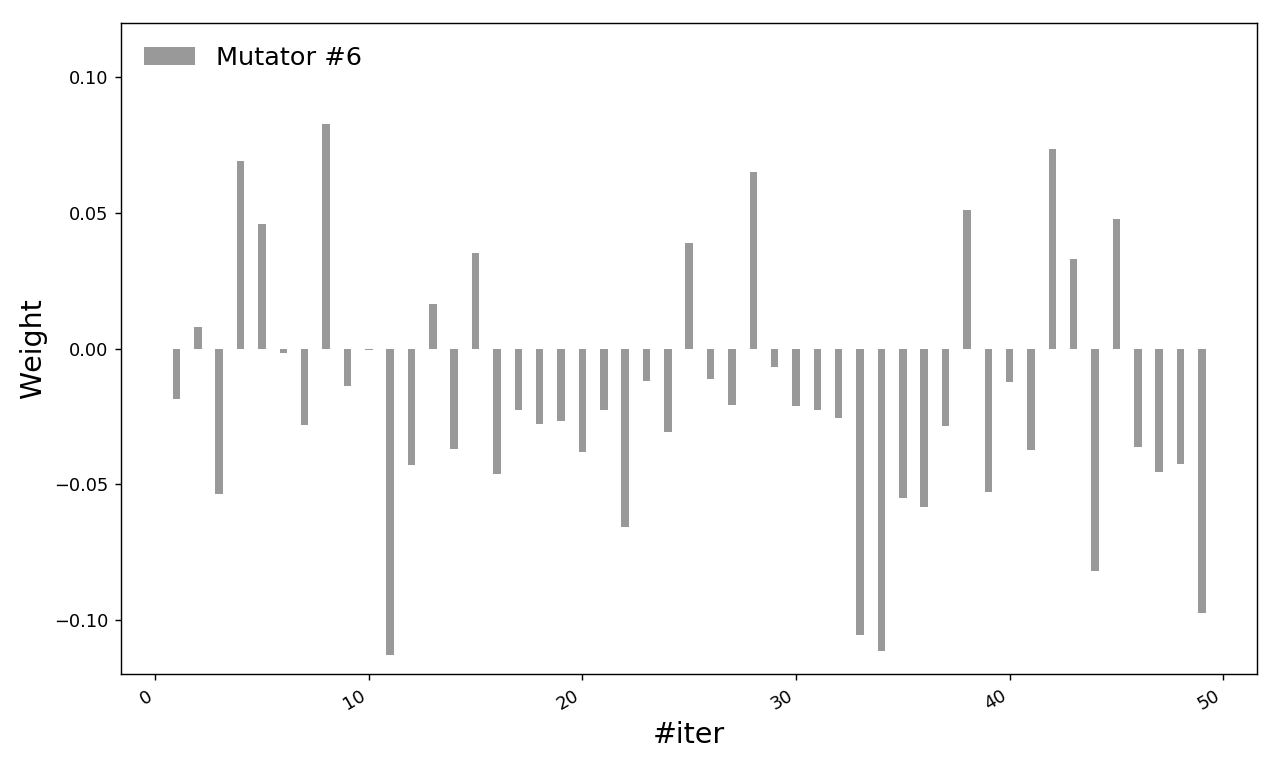}
  } 
  \subfigure[Mutator \#7]{ 
    \label{diff:g}
    \includegraphics[width=4cm]{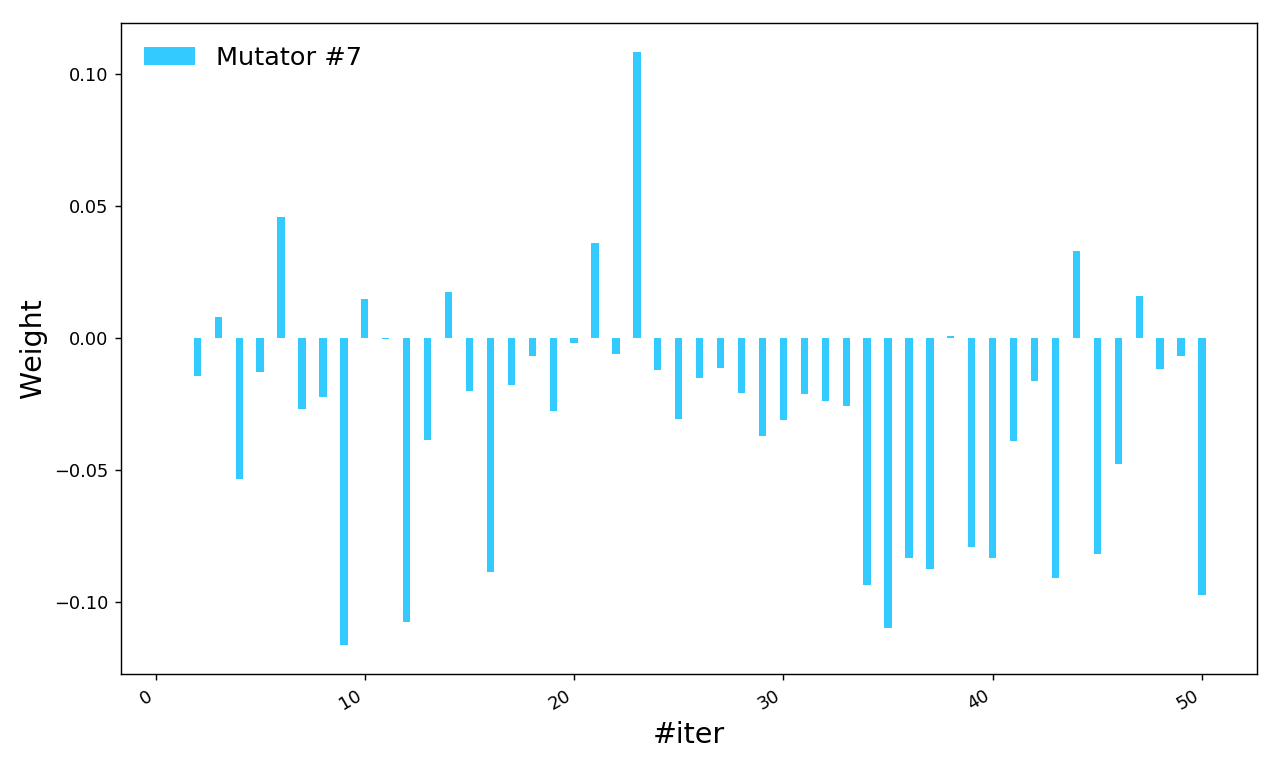}
  } 
  \subfigure[Mutator \#8]{ 
    \label{diff:h}
    \includegraphics[width=4cm]{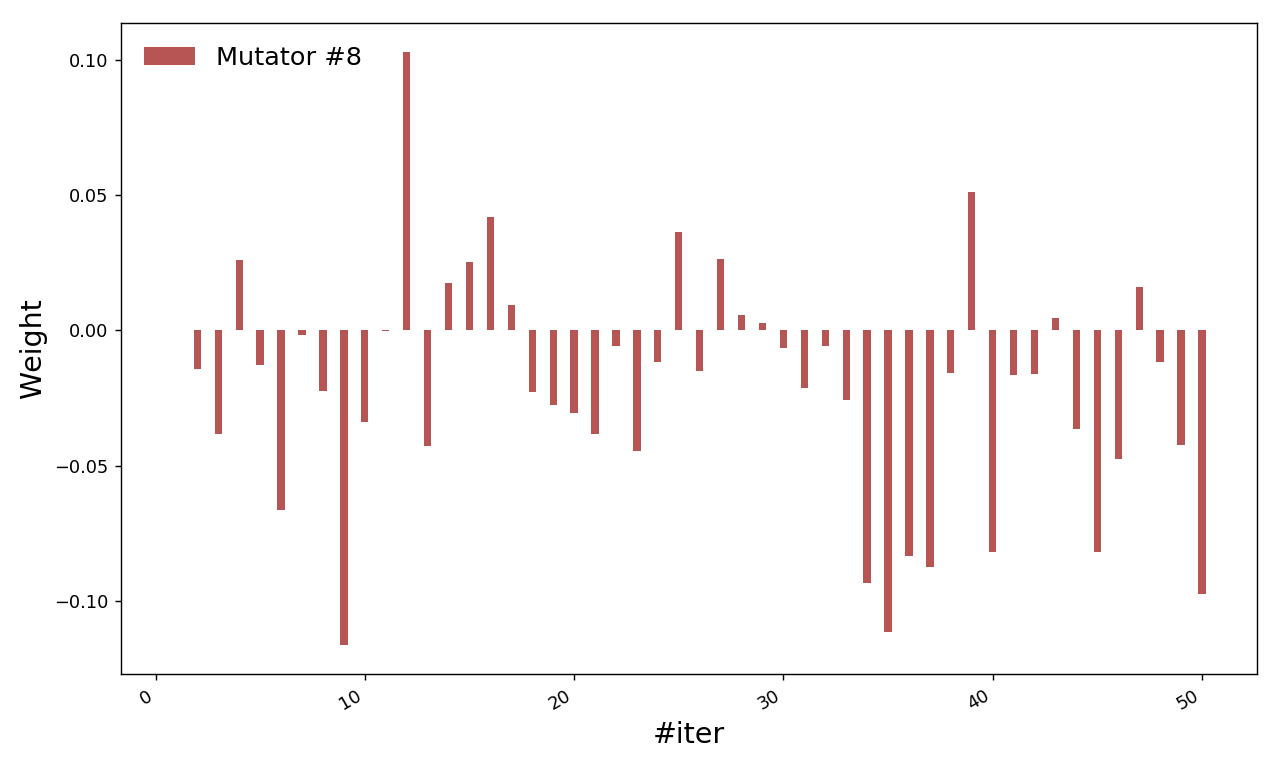}
  } 
  \caption{Weight changing of each mutator recorded in each mutation iteration.} 
  \label{Diff}
\end{figure*}

\begin{table*}[htp]
\caption{Final weights of each mutator}
\begin{center}
\begin{tabular}{ | c | c | c | c | }
  \hline
  \makebox[20mm][c]{\textbf{Mutator}} & \makebox[40mm][c]{\textbf{Object}} & \makebox[70mm][c]{\textbf{Method}} & \makebox[20mm][c]{\textbf{Weight}} \\ \hline
  \#2 & conditional operator & Modify all conditional operators & 0.2703 \\ \hline
  \#3 & arithmetic operator & Modify all arithmetic operators & 0.1691 \\ \hline
  \#5 & loop operator & Increase the iteration of loop by 99 & 0.1180 \\ \hline
  \#8 & control structure & Insert continue or break statement & 0.1043 \\ \hline
  \#1 & local variable & Modify all variable attributes & 0.1022 \\ \hline
  \#6 & assert statement & Delete or insert an assert statement & 0.0963 \\ \hline
  \#4 & function property & Modify all function attributes & 0.0765 \\ \hline
  \#7 & return statement & Delete the return statement & 0.0631 \\ \hline
\end{tabular}
\label{weight}
\end{center}
\end{table*}

\begin{figure}
    \centering
    \includegraphics[width=8cm]{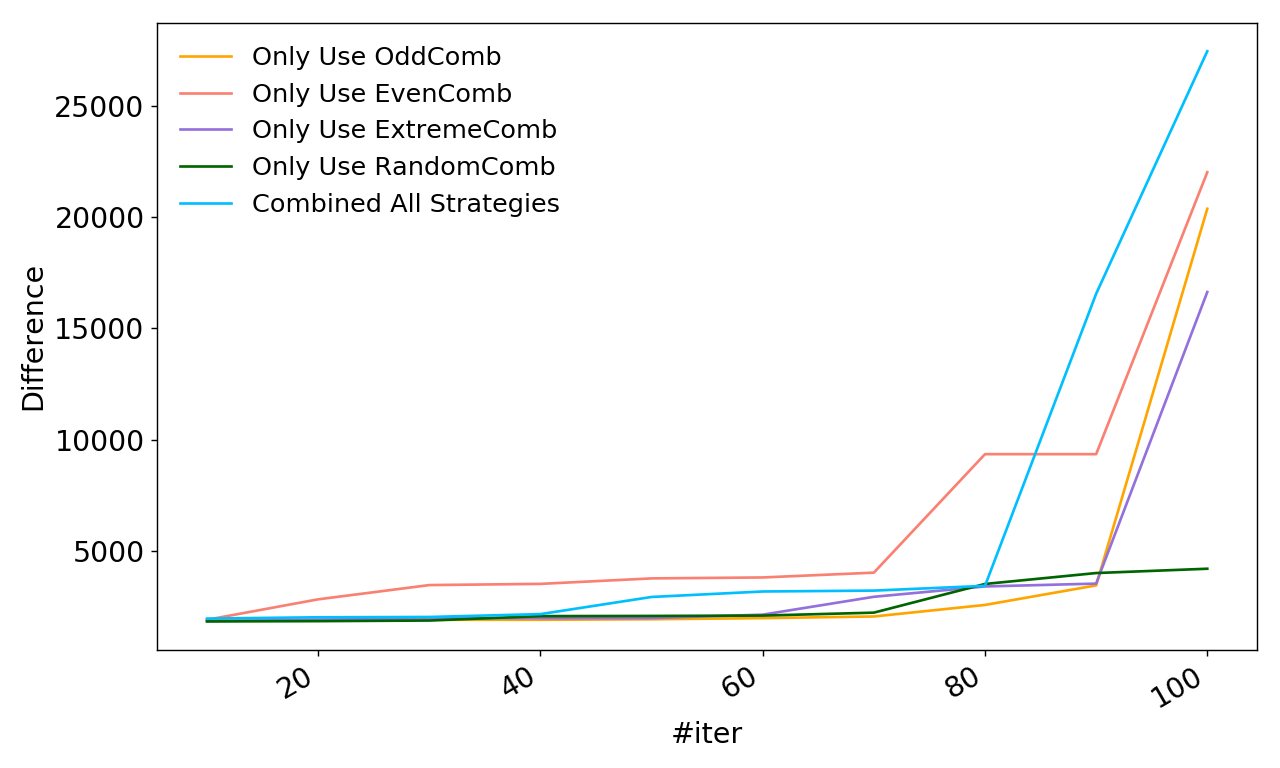}
    \caption{Efficiency of different mutation strategies.}
    \label{Diff-1}
\end{figure}

As mentioned in \S \ref{ContractMutation}, we designed 5 strategies for mutator combination, and their performance statistics are presented in  Fig. \ref{Diff-1}. At the beginning of mutation, EvenComb strategy had the fastest rising rate. Within 30 iterations, the metric difference $diff$ increased from 346 to 1,840. However, it then had the difficulty in jumping out of the local optimal solution and grew slowly. After the 80th iteration, AllComb strategy found a new breakthrough and began to grow rapidly. Eventually, it got the largest metric difference. It also shows that the randomly choose different strategies with guiding information in each iteration can generate high-quality seeds most efficiently. As for RandomComb strategy without weight guiding information, which is completely based on randomness, had the worst performance in rising rate and final value. Those statistics indicate that the weight information of each mutator has a certain guiding significance and facilitates the combined mutation strategy. 

To investigate the effects of each mutator and the corresponding weight, we analyze the procedure of AllComb strategy in detail. 
Firstly, we generate a tuple of input parameters based on the function signature. Then, with the function signature and input parameters unchanged, different strategies are randomly adopted to mutate the original contract for 50 iterations. After that, the feedback metric difference and the corresponding mutator weights in each iteration are recorded, shown in Fig. \ref{Diff}. The mutator ID matches the first column of Table \ref{weight}, which shows the mutator weights statistics after mutating 36,295 contracts. There are eight mutators involved, with an average weight of 0.125, so we use this average value as a baseline to observe whether each mutator plays a positive or negative role in each iteration of AllComb strategy.

In Fig. \ref{Diff}(b)(c), most regions are concentrated above the baseline, and the peak value even reaches to 0.6. These two mutators are respectively used for arithmetic operators and conditional operators modification. For the arithmetic operators, mutator changes the results of computation and affects the subsequent execution path; and for the conditional operators, mutator changes the executed module, involves different execution opcodes and causes the increasing of difference. On the contrary, in Fig. \ref{Diff}(g), most areas are concentrated below the baseline, indicating that this mutator didn't play a leading role in most iterations. The reason is that deleting the return value of a function is an irreversible operator and will not  have effect after one modification. The remaining five mutators' weights are mainly near the baseline ($\pm 0.05$). Among them, \#1 and \#5 have a higher peak value, which shows that they are easier to trigger the trend of rising under some corner cases; meanwhile, \#4, \#6 and \#8 have little influence on the increasing of difference. 
\begin{table*}[!htp]
\caption{Description of 5 high-risk vulnerabilities detected by EVMFuzz}
\vspace{-0.5cm}
\begin{center}
\begin{tabular}{ | c | c | c | c | m{70mm} | c | }
  \hline
  \textbf{CVE-ID} & \textbf{platform} & \textbf{version} & \textbf{language} & \quad \quad \quad \quad \quad \quad \quad \quad \tabincell{c}{\textbf{description}} & \textbf{created date} \\ \hline
  CVE-2018-18920 & Py-EVM & v0.2.0-alpha.33 & python & Py-EVM v0.2.0-alpha.33 allows attackers to make a vm.execute\_bytecode call that triggers illegal values shown in stack. & 20181103 \\ \hline
  CVE-2018-19183 & js-evm & v2.4.0 & JavaScript & ethereumjs-vm 2.4.0 allows attackers to cause a denial of service (vm.runCode failure and REVERT) via a "code: Buffer.from(my\_code, 'hex')" attribute. & 20181111 \\ \hline
  CVE-2018-19184 & geth & v1.8.17 & golang & cmd/evm/runner.go in Go Ethereum (aka geth) 1.8.17 allows attackers to cause a denial of service (SEGV) via crafted bytecode. & 20181111 \\ \hline
  CVE-2018-19330 & aleth & v1.5.0-alpha.6 & cpp & ** RESERVED ** Details would be public after the vulnerability has been repaired to avoid potential attack. & 20181117 \\ \hline
  CVE-2019-7710 & aleth & v1.5.0-alpha.7 & cpp & ** RESERVED ** Details would be public after the vulnerability has been repaired to avoid potential attack. & 20190210 \\ \hline
\end{tabular}
\label{cve}
\end{center}
\end{table*}
From the statistics, it is reasonable to conclude that EVMFuzz could efficiently generate corner cases to trigger different outputs among EVMs with guiding information of $gasUsed$ and opcode sequence.

\subsection{Is it possible to find EVM bugs through differential fuzz  testing?}
We have demonstrated that EVMFuzz can efficiently generate seed contracts and perform differential fuzz testing to find difference across different EVM platforms. After discovering thousands of output inconsistencies, we conducted the manual analysis and tried to explore the root causes. 

We ensured its reproducibility and then carefully reviewed the source code of EVMs. Finally, we found defects in the EVM platforms, of which, 5 vulnerabilities were registered as Common Vulnerabilities and Exposures, numbered as CVE-2018-18920, CVE-2018-19183, CVE-2018-19184, CVE-2018-19330 and CVE-2019-7710, shown in Table \ref{cve}. 
Except for these 5 EVM vulnerabilities, there were only 6 existing vulnerabilities associated with EVM among 112,913 CVE entries~\cite{cveEntry}, while smart contract vulnerabilities were over 500. We select two previous unknown CVE instances for a specific explanation.



\begin{figure}[!htbp]
\centering	
\includegraphics[width=0.45\textwidth]{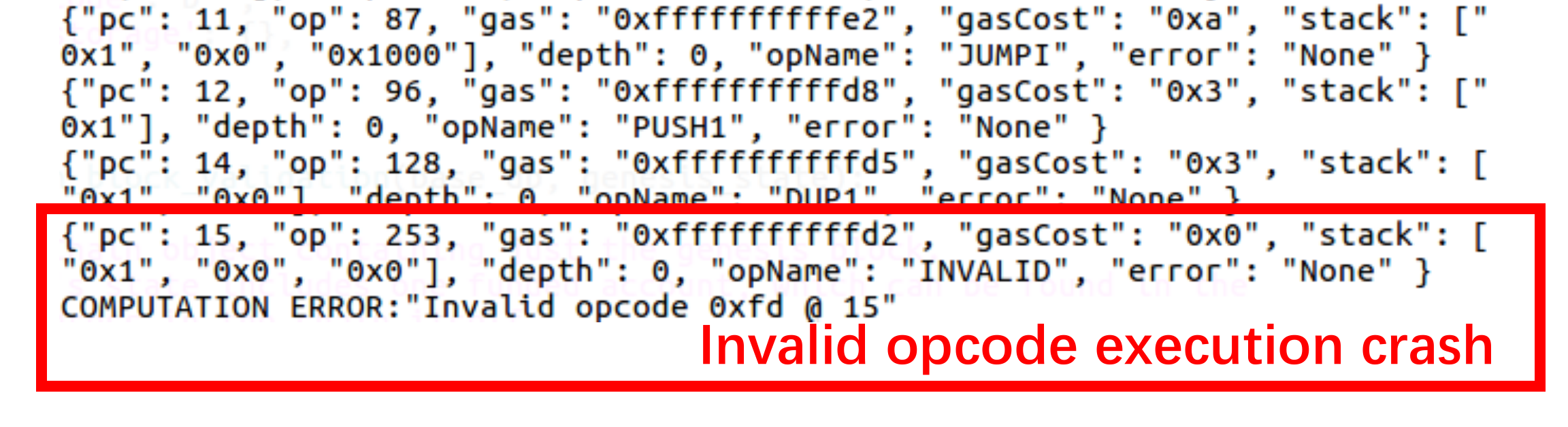}
\vspace{-0.5cm}
\caption{Invalid execution output of Py-EVM.}
\label{py-cve-output}
\end{figure}

%
\vspace{0.2cm}
\noindent\textbf{\textsl{CVE-2018-18920 Case Study.\quad}} This is a runtime error on Py-EVM. Fig. \ref{py-cve-output} shows a contract Origin with one construct function.

Function $vm.execute\_bytecode$ in Py-EVM takes contract bytecode and data as input and executes the bytecode along with the data. When we used this function to execute the Origin's bytecode, with the parameter value,  
the execution terminated abnormally. 

The execution result is shown in Fig. \ref{py-cve-output}.  This issue has been included in CVE database as CVE-2018-18920, it describes that this vulnerability would result in an execution failure because of an invalid opcode, and smart contracts can be executed indefinitely without gas being paid. 
Once this vulnerability is exploited, the attackers can make the Ethereum network abnormally congested without cost, which will cause the Ethereum ecology to be imbalanced.

\lstset{language=Solidity}
\label{sol:3}
\begin{lstlisting}
pragma solidity ^0.4.24;
contract Origin {
  address public owner;
  /**
   * @dev The Ownable constructor sets the original `owner`
   * of the contract to the sender contract.
   */
  function Origin() {
    owner = msg.sender;} 
}
\end{lstlisting}

\vspace{0.2cm}
\noindent\textbf{\textsl{CVE-2018-19184 Case Study.\quad}} This is an execution segmentation violation that occurred on EVM of Go Ethereum (geth).  
The code associated with this vulnerability was in the cmd/evm folder, where the exception handling mechanism of EVM before geth v1.8.14 did not cover enough corner cases. 
Although the problematic code snippet is not the API that directly exposed to the end users, this problem can be exploited by malicious attackers to cause the Ethereum platform denial of service. 
\vspace{-0.2cm}
\begin{figure}[!htbp]
    \centering
    \includegraphics[width=8.3cm]{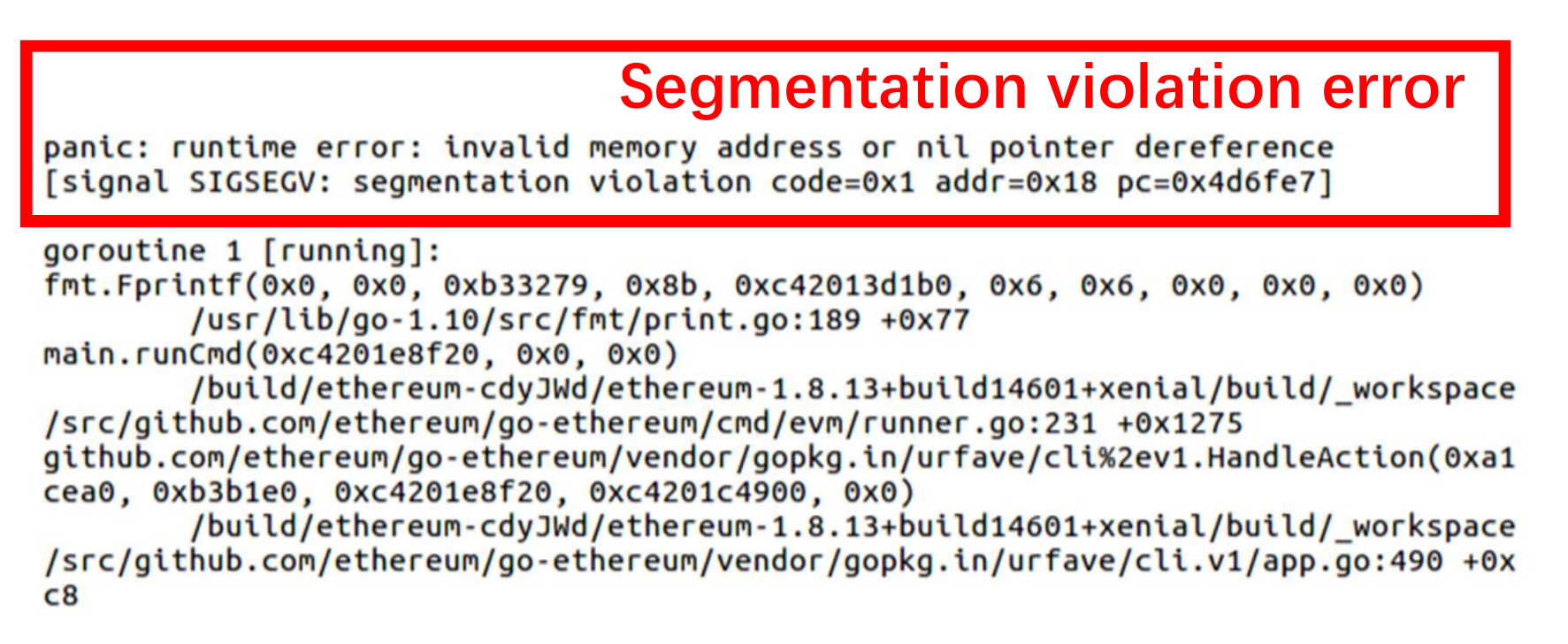}
    \vspace{-0.5cm}
    \caption{Segmentation violation error on geth-evm v1.8.13.}
    \label{geth-CVE}
\end{figure}

\section{Discussion}
\label{Discussion}
We propose a differential fuzz testing framework and have found some bugs, but there are still some deficiencies that need to be improved in our future work.

\begin{enumerate}[leftmargin=*]
\item \textbf{More mutators design.}
Considering the correctness of compiling and the logical complexity, we designed eight mutators, whose quantity still needs improving. Currently, we just modify the key attribute values based on CAST. In the future, we will make more semantic level mutations based on source code, such as adding internal function calls or contract inheritance relations. In the future, we can even automatically generate contracts for testing, similar to the the C program generation tool Csmith \cite{yang2011finding} and the SQL generation tool SQLsmith \cite{kerstensqalpel}. 

\item \textbf{More efficient priority scheduling.}
In order to maintain the priority queue of mutator and candidate seeds, we adopt the data structure of $PriorityQueue$, which is implemented by $heap$. The insertion and deletion operations take the complexity of $\Theta$(logn), and the time complexity of sorting is $\Theta$(nlogn). Now, the scale of mutators is small and the length of candidate seeds is short, so the time overhead is small, but the storage structure still needs to be optimized to further improve efficiency. Furthermore, the mutation priority is now based on the sum of two parts, the first part is difference priority and the second part is time priority. We can introduce coefficients $a$ and $b$ to balance these two parts in the future. For example, the difference priority might account for more proportion than the time priority. 

\item \textbf{Support of more EVM implementations.} 
The core idea of differential testing is testing on at least two objects with the same functions. Therefore, we analyze the differences in four EVM platforms: js-evm, Py-EVM, geth and aleth. However, in the real world, many platforms based on other programming languages are also widely used, and they may also have some fatal vulnerabilities in implementation. We will further incorporate them into our platform. 

\item \textbf{More accurate selection metrics.}
In the selection of mutated seed contracts, we use $gasUsed$ and opcode sequence difference as the inducer for triggering EVM vulnerabilities. But there are still some problems, such as insufficient quantity and the lack of unity for data types, waiting for further improvement. Furthermore, even for the two difference indicators, we can also introduce coefficients $a$ and $b$ to balance the $gasUsed$ and the opcode sequence difference in the future. For example, the weight of the opcode sequence difference might account more than that of $gasUsed$ difference. We can even dynamically adjust the coefficients during the whole fuzzing stages.  

\end{enumerate}

\vspace{-0.1cm}
\section{Related Work}
\label{Related-work}

\noindent \textbf{Fuzzing Technique.} \quad
Fuzzing is an automatic testing technique that covers numerous boundary cases using invalid data as application input to better ensure the absence of exploitable vulnerabilities~\cite{Liang2018FuzzingSO}.
Some popular AFL~\cite{AFL} family tools~\cite{Bhme2016CoveragebasedGF,Bhme2017DirectedGF,Lemieux2018FairFuzzAT,Stephens2016DrillerAF,Wang2018SAFLIA,Liang2018PAFLEF,Schumilo2017kAFLHF,TriforceAFL,WinAFL,Chen2018EnFuzzFE} apply various strategies to boost fuzzing process, including symbolic execution, schedule algorithm and so on. 
For example, EnFuzz~\cite{Chen2018EnFuzzFE} integrates multiple fuzzing strategies to obtain better performance and generalization ability than that of any constituent fuzzer alone.
There are also some tools focus on fuzzing in other domains, for example, 
QuanFuzz~\cite{Wang2018QuanFuzzFT} is a search-based test input generator for the quantum programs.

\vspace{0.1cm}
\noindent \textbf{Differential Testing.} \quad
Differential testing~\cite{McKeeman1998DifferentialTF} has been very successful in uncovering differences between independent implementations with similar intended functionality.
For example, Chen \etal's perform differential testing of JVMs using MCMC sampling for input generation~\cite{Chen2016CoveragedirectedDT}. 
Differential testing has also been used for deep learning systems. DeepXplore~\cite{Pei2017DeepXploreAW} was presented as the state-of-the-art white-box differential testing framework for deep learning systems and first introduce the concept of neuron coverage as a testing metric. DLFuzz~\cite{Guo2018DLFuzzDF} extends differential testing framework for DL systems with the comparisons of multiple similar inputs, and does not need multiple platforms. 

\vspace{0.1cm}
\noindent \textbf{Smart Contract Validation.} \quad
Smart contracts have been shown to be exposed to severe vulnerabilities~\cite{Atzei2016ASO,hirai2016formal}, and many efforts have been devoted to ensure the correctness. 
Luu \etal ~\cite{Luu2016MakingSC} designed Oyente, which builds the control-flow graph from the bytecode and then performs symbolic execution and checks whether there exist any vulnerable patterns.
Nikolic \etal ~\cite{Nikolic2018FindingTG} designed MAIAN for reasoning about tracing properties to detect vulnerable smart contracts. It specified three typical smart contracts vulnerabilities based on trace properties. Zeus~\cite{Kalra2018ZEUSAS} is a sound analyzer that translates smart contracts to the LLVM framework and uses XACML as a language to write properties. 

\vspace{0.1cm}
\noindent\textbf{EVM Semantics Formalization.} \quad
Semantics formalization on EVM tries to tackle a complex mix between requirements for high assurance and a rich adversarial model of Ethereum. KEVM~\cite{KEVM} is the first fully executable formal semantics of EVM, which is created based on a framework of executable semantics named the K framework. A Lem~\cite{Mulligan2014Lem} language based EVM implementation provides a formal specification of the interface between the smart contract execution and the execution environment. 
This EVM definition can be used to prove invariant and safety properties of smart contracts deployed on Ethereum.

\vspace{0.1cm}
\noindent \textbf{Main Difference.}  \quad
Different from the above work, EVMFuzz mainly focuses on discovering the defects and vulnerabilities in EVM. 
In the case that most people are concerned about smart contracts validation, it is necessary to focus our attention on the validation of Ethereum infrastructure--EVM. 
Particularly, EVMFuzz combines the basic ideas of fuzzing and takes advantage of EVMs' multi-implementation to quickly find output discrepancies and reduce manual checks. 
Within EVMFuzz, we also define the domain specific EVM test indicators to guide the differential fuzzing process with different contract mutation and selection strategies.

\vspace{-0.1cm}
\section{Conclusion}
\label{Conclusion}
In this paper, We propose EVMFuzz, the first automated differential fuzz testing framework, to efficiently find vulnerabilities of EVMs implemented by different programming languages. We introduce the definition of EVM fuzz testing metrics--$gasUsed$ and opcode sequence, which measure the internal difference in execution information between EVMs. We design 8 mutators for smart contracts, so that EVMFuzz can generate plenty of seed contracts without syntax error in a short time. EVMFuzz uses the metric difference as a guidance for seed contract preserving and employs a dynamic priority scheduling algorithm for selecting the contract in the next iteration. We evaluate EVMFuzz based on numerous mutation on 36,295 real-world smart contracts. Among the generated 253,153 smart contracts, more than half successfully showed the differential performance of $gasUsed$ and opcode indicator, including 1,596 variant contracts triggered inconsistent output results among the four EVM platforms. With manual root cause analysis, 5 vulnerabilities have been assigned with unique CVE IDs. Our future work mainly includes developing more general smart contract mutators, generators and priority scheduling algorithms, and conducting more extensive evaluations on more EVMs.

\bibliographystyle{ACM-Reference-Format}
\bibliography{CCS19}

\end{document}